\documentclass[journal,10pt]{IEEEtran}
\usepackage[utf8]{inputenc}
\usepackage{graphicx} 
\usepackage{booktabs} 
\usepackage{amsmath}
\usepackage{amsthm}
\usepackage{bm}
\usepackage{gensymb}
\usepackage{physics}
\usepackage{mathtools, nccmath}
\usepackage{algorithm}
\usepackage{algpseudocode}
\usepackage[table]{xcolor}
\usepackage{epstopdf}
\usepackage{balance}
\usepackage{setspace}
\usepackage{amsfonts}
\usepackage{amssymb}
\usepackage{comment}
\usepackage{romannum}
\usepackage[justification=justified,skip=0pt]{caption}
\usepackage{bbm}
\usepackage{cite}
\usepackage{adjustbox}
\usepackage[english]{babel}
\usepackage{makecell}

\theoremstyle{definition}

\usepackage{multicol}

\usepackage{mathrsfs}
\usepackage{tikz}
\usepackage{forest}
\usetikzlibrary{shadows}
\usepackage{hhline}
\usepackage{bookmark}
\usepackage{multirow}

\title{Unsourced Random Access:\\A Comprehensive Survey}

\author{Mert Ozates, Mohammad Javad Ahmadi, Mohammad Kazemi, Deniz G\"und\"uz and Tolga M. Duman
\thanks{Mert Ozates is with the System Architectures Department, IHP - Leibniz Institute for High-Performance Microelectronics, 15236 Frankfurt (Oder), Germany (e-mail: oezates@ihp-microelectronics.com).}
\thanks{Mohammad Javad Ahmadi is with the Chair of Information Theory and Machine Learning, Technische Universität Dresden, 01062
Dresden, German (E-mail: mohammad\_javad.ahmadi@tu-dresden.de)}
\thanks{Mohammad Kazemi and Deniz G\"und\"uz are with the Department of Electrical and Electronic Engineering, Imperial College London, London SW7 2BT, U.K. (e-mail: \{mohammad.kazemi,d.gunduz\}@imperial.ac.uk)}
\thanks{Tolga M. Duman is with the Department of Electrical and Electronics Engineering, Bilkent University, Ankara, 06800, Turkey (email: duman@ee.bilkent.edu.tr)}%
\thanks{This work was supported by the Scientific and Technological Research Council of Turkey (TUBITAK) under Grant 119E589. Mert Ozates' work is supported by the Federal Ministry of Education and Research of Germany in the programme of “Souverän. Digital. Vernetzt” joint Project 6G-RIC, project identification number: 16KISK026. Mohammad Kazemi’s work was funded by UK Research and Innovation (UKRI) under the UK government’s Horizon Europe funding guarantee [grant number 101103430].}
}

\begin{document}
\pagenumbering{arabic}
\maketitle

\begin{abstract}
Multiple access communication systems enable numerous users to share common communication resources, playing a crucial role in wireless networks. With the emergence of the sixth generation (6G) and beyond communication networks, supporting massive machine-type communications with sporadic activity patterns is expected to become a critical challenge. Unsourced random access (URA) has emerged as a promising paradigm to address this challenge by decoupling user identification from data transmission through the use of a common codebook. This survey offers a comprehensive overview of URA solutions, encompassing both theoretical foundations and practical applications. We present a systematic classification of URA solutions across three primary channel models: Gaussian multiple access channels (GMACs), single-antenna fading channels, and multiple-input multiple-output (MIMO) fading channels. For each category, we analyze and compare state-of-the-art solutions in terms of performance, complexity, and practical feasibility. Additionally, we discuss critical challenges such as interference management, computational complexity, and synchronization. The survey concludes with promising future research directions and potential methods to address existing limitations, providing a roadmap for researchers and practitioners in this rapidly evolving field.

\end{abstract}

\begin{IEEEkeywords}
Multiple access, random access techniques, grant-free, unsourced random access, machine-type communications, Internet of things. 
\end{IEEEkeywords}

\section{Introduction}


The rapid proliferation of Internet of Things (IoT) devices is driving a fundamental shift in wireless communication systems. Current estimates project over 38 billion IoT connections by 2029, representing a more than twofold increase compared to the 15 million devices in 2023 \cite{Ericsson2024}.
One example of such systems is machine-type communications (MTC) (also known as machine-to-machine (M2M) communications), which, as a key aspect of the IoT paradigm, enables machines, sensors, and devices to exchange information without direct human intervention. MTC plays a crucial role in applications ranging from industrial automation and smart cities to healthcare monitoring and environmental sensing \cite{Shariat2017From}. Specifically, massive MTC is anticipated to accommodate a density of up to one million devices (per square kilometer), which typically operate with low computational and storage capabilities, low duty cycles, and small payloads \cite{Dutkiewicz2017Massive}. 
This massive scale presents unprecedented challenges for communication systems as traditional wireless access paradigms have been designed for a relatively small number of high-data-rate users rather than an enormous number of devices with sporadic low-rate transmissions.

Multiple access (MA) techniques that enable different users to share a common communication medium, have been fundamental to wireless communication systems \cite{Biglieri2007}. MA technologies can be broadly categorized into two categories: centralized access and random access (RA). In centralized access (also known as \textit{coordinated access}), a central authority controls and manages access to the shared communication channel by allocating distinct resources to each user for data transmission. 
This approach has been the predominant MA mode in the fifth generation (5G) and previous cellular technologies. 
However, centralized coordination becomes impractical for massive IoT scenarios due to prohibitive signaling overhead and coordination complexity. In contrast, in RA (also known as \textit{uncoordinated access}), users transmit their messages without requiring explicit coordination with a central authority. 
This approach reduces signaling overhead and, in some cases, latency. These advantages make RA solutions particularly suitable for next-generation (6G and beyond) communication networks, especially when serving a large number of users \cite{OzatesThesis,AhmadiThesis}.

 \begin{figure*}[t]
	\centering
\includegraphics[width=.7\linewidth]{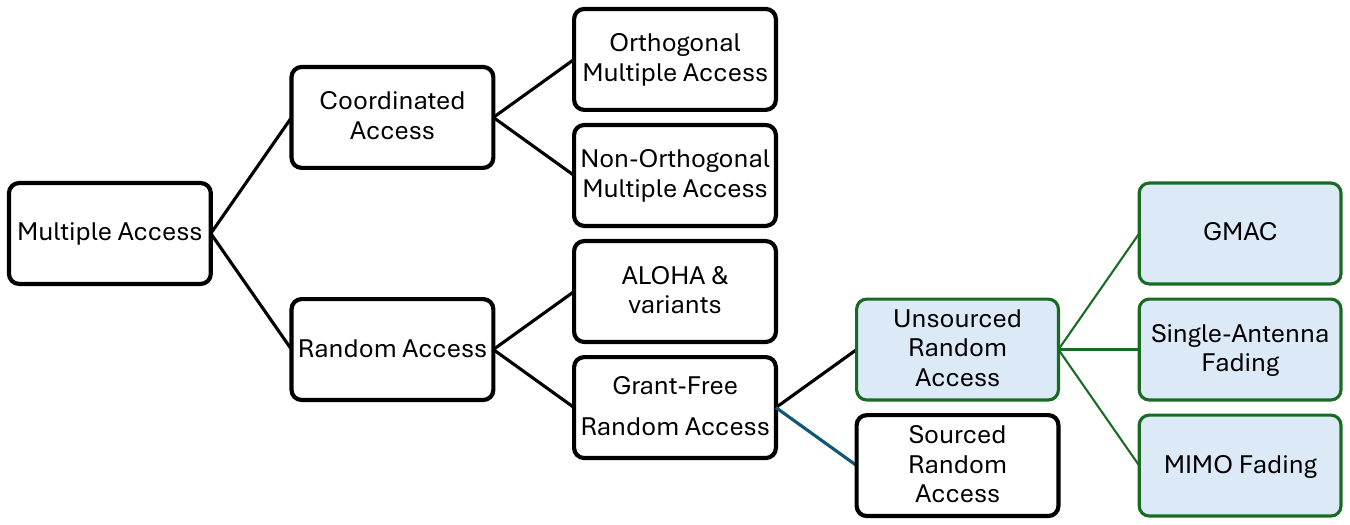}
		\caption{ Taxonomy of different MA techniques.}	\label{FigFlowChart}
\end{figure*}

Traditional RA solutions, such as ALOHA \cite{Abramson1970The} and its variants, suffer from an excessive number of collisions as the number of users increases. More sophisticated grant-free RA schemes \cite{Choi2022Grant} employ unique signatures or preambles for each potential user to facilitate collision reduction at the physical layer. Grant-free RA solutions can accommodate more users compared to ALOHA, as they can accomplish user scheduling through a limited amount of feedback \cite{Mahmood2019Sign}, or by exploiting device-specific patterns such as codebook structures and spreading sequences \cite{Mahmood2019Uplink}. 
\textcolor{black}{On the other hand, with the number of devices within a certain coverage area reaching millions \cite{survey}, it is not feasible to assign a distinct pilot/preamble to each device, as this would result in prohibitive receiver complexity. Therefore, existing grant-free RA solutions cannot address the multiple access problem in emerging massive access networks.}



To mitigate the challenge of identifying and decoding messages from such a large number of potential devices, Polyanskiy introduced the {\it unsourced random access} (URA) paradigm in \cite{polyanskiy2017perspective}. \textcolor{black}{URA relies on the observation that, despite their increasing numbers, MTC and IoT devices are activated only sporadically; that is, at any time instant, only a handful of devices are active simultaneously.} URA fundamentally reimagines the massive access problem by decoupling user identification from data transmission. Instead of assigning unique preambles or codebooks, users employ a common codebook, and the receiver focuses solely on recovering the list of transmitted messages. \textcolor{black}{Since a common codebook is employed, the user identity is eliminated and the system can operate irrespective of the total number of devices, where only the number of active devices becomes important.} This paradigm shift drastically reduces system complexity, and enables practical implementations for truly massive numbers of devices. Following Polyanskiy's pioneering work, several low-complexity coding solutions have been developed in the subsequent literature, where successive interference cancellation (SIC) techniques are employed to resolve collisions up to a certain level, depending on the user's transmit powers. Many techniques, such as random spreading \cite{Gkagkos2023FASURA}, multistage orthogonal pilots \cite{Ahmadi2023Unsourced}, and time frame slotting \cite{Ozates2023Aslotted} have been employed to mitigate collisions and interference. While a single-antenna receiver in a Gaussian MA channel (GMAC) setup is considered in the earlier works,  the literature has shifted towards fading channels with multiple-antenna receivers, especially massive multiple-input multiple-output (MIMO), to support more practical channel models and an even greater number of users. A block diagram of different MA techniques is depicted in Fig. \ref{FigFlowChart} {\color{black}alongside the three main setups in URA research, namely, GMAC, and single-antenna and MIMO fading channels. 
}

 \begin{table}
\footnotesize
\centering
\caption{\textcolor{black}{Summary of the surveys on URA.}}
\begin{tabular}{|m{.18\linewidth}|m{.7\linewidth}|}
\hline
\centering \cellcolor[HTML]{BDBDBD}\textcolor{black}{Paper}   &  \cellcolor[HTML]{BDBDBD}\textcolor{black}{Scope}\\
\hline
\makecell{ \textcolor{black}{Wu et al., } \\ \textcolor{black}{2020 \cite{survey}}} & \textcolor{black}{ Review of many-user MAC and URA with a focus on GMAC.}\\ \hline
\makecell{ \textcolor{black}{Li et al., } \\ \textcolor{black}{2022 \cite{Li2022Unsourced}}}& \textcolor{black}{Review of basics of URA and coding schemes on URA over GMAC. }\\\hline
\makecell{ \textcolor{black}{Gao et al.,} \\ \textcolor{black}{2024 \cite{Gao2024survey}}  }  & \textcolor{black}{  Review of grant-free massive access with a focus on compressed sensing-based schemes.} \\\hline
\makecell{ \textcolor{black}{Liva et al., } \\ \textcolor{black}{2024 \cite{Liva2024}}}& \textcolor{black}{ Review of URA over GMAC and showing that it can support 3GPP standardization.}\\\hline
\makecell{ \textcolor{black}{This survey}}  & \textcolor{black}{A comprehensive review of the URA literature, covering all existing solutions in GMAC and single-antenna and MIMO fading channels.}\\\hline
\end{tabular}
\label{Table_Surveys}
\normalsize
\end{table}

\subsection{\textcolor{black}{Objectives and Contributions}}

\textcolor{black}{ \textcolor{black}{Over the years since Polyanskiy's initial work, a large number of papers have appeared studying URA from both theoretical and practical perspectives. Although prior efforts have been made to review existing works on URA, our survey fills an important gap by providing a systematic classification of the literature and covering more recent works on practical channel models and device characteristics.} The first survey on massive access was \cite{survey}, in which the authors review many-user MA, where each user is assigned an individual codebook (i.e., the coordinated approach), as well as URA, which they call uncoordinated massive access. However, since URA was in its early stages at that time, the survey in \cite{survey} covers mainly the earlier works over GMAC. In \cite{Li2022Unsourced}, the focus is again on URA over GMAC, and the authors review the existing design paradigms and coding schemes for URA over GMAC, while URA over fading channels and in MIMO systems are presented as open problems. In a less URA-oriented survey \cite{Gao2024survey}, an overview of compressive sensing-based grant-free massive access is provided, which also includes some compressed sensing-based URA schemes. In a recent survey \cite{Liva2024}, through an overview of the existing URA schemes studying mostly GMAC, the authors present URA as a suitable candidate for next-generation communication systems and highlight that it can support the 3rd Generation Partnership Project (3GPP) standardization efforts.}

\begin{figure}[t]
    \centering
    \includegraphics[trim={0cm 0cm 0cm 0cm},clip,width=1\linewidth]{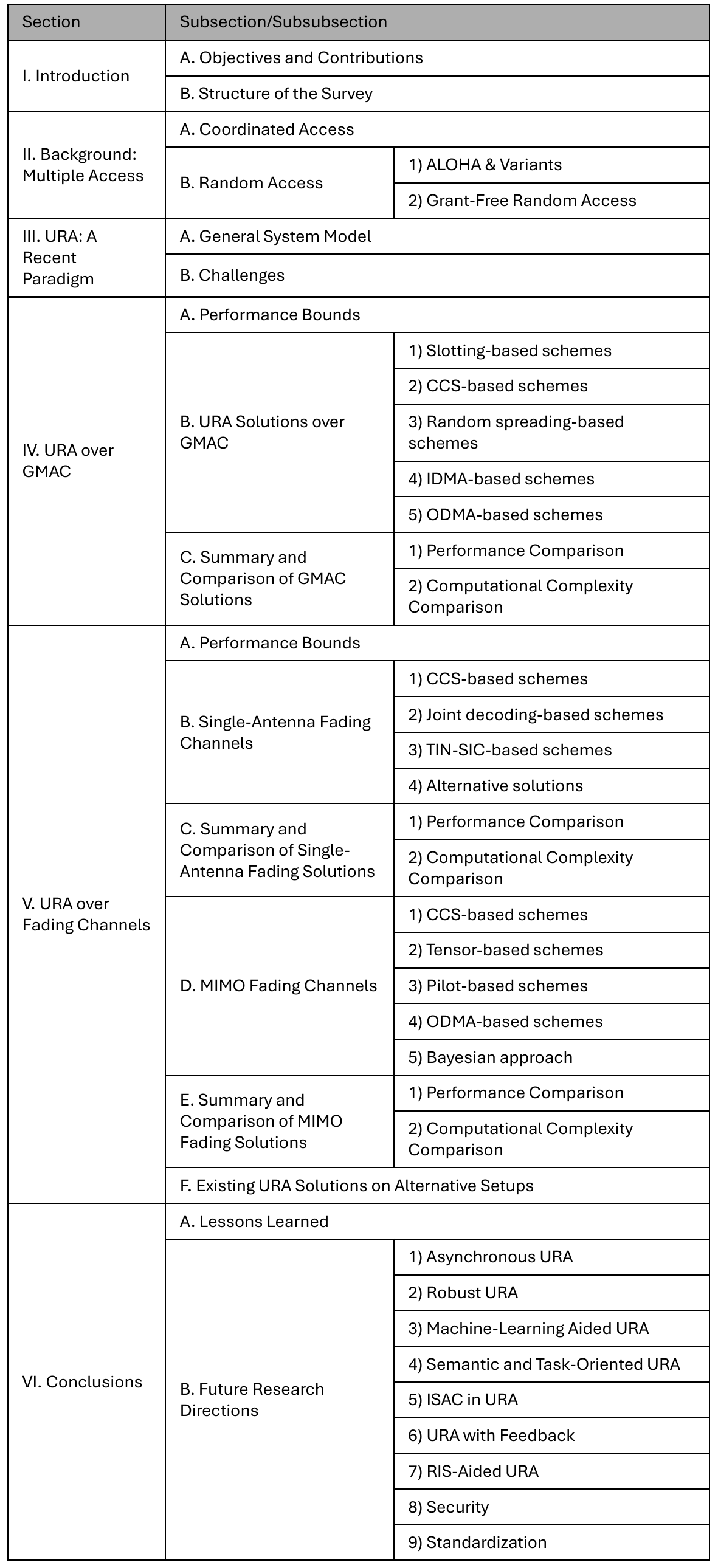}
    \caption{\textcolor{black}{Structure of the survey.}}
    \label{structure}
\end{figure}

\textcolor{black}{
While the existing surveys focus on the GMAC setup, we also cover the URA solutions that take into account channel fading, which comprise the majority of recent works in the URA literature. More specifically, we conduct a comprehensive review of the existing URA solutions in the literature across three main setups, namely, URA over GMAC, and single-antenna and MIMO fading channels, accompanied by a discussion on the main challenges, possible solutions, and promising research directions. We study works that propose coding schemes, as well as those that provide fundamental information-theoretical analysis for URA. To the best of our knowledge, this is the first comprehensive survey on URA, reviewing this crucial emerging paradigm for 6G and beyond communication networks in all aspects. A summary of the existing surveys and our contribution can be found in Table \ref{Table_Surveys}.}

\textcolor{black}{In particular, the paper contains the following key contributions:
\begin{itemize}
    \item Comprehensive overview: Delivering an extensive survey of URA solutions, examining both theoretical and practical perspectives.
    \item Classification: Systematically categorizing URA solutions according to the fundamental channel models: GMAC, single-antenna fading channels, and MIMO fading channels.
    \item Performance and feasibility analysis: Evaluating and comparing state-of-the-art URA solutions in terms of energy efficiency, computational complexity, and practical viability.
    \item Identifying challenges: Addressing essential challenges in URA in terms of interference, collisions, channel estimation, and computational complexity.
    \item Guiding future research: Proposing promising future research directions and potential solutions to overcome existing limitations in URA systems.
\end{itemize}
}

\textcolor{black}{Throughout the survey, we adopt a unifying narrative: to demonstrate how URA techniques progress from theoretical achievability results on the GMAC to increasingly practical frameworks under fading and MIMO channels. This progression underscores the significance of URA as an enabler of massive connectivity in emerging 6G and IoT systems.}

\begin{table*}[t]
\footnotesize
\centering
\caption{List of acronyms}
\begin{tabular}[t]{|p{0.12\linewidth}|p{0.31\linewidth}|} 
 \hline
 \cellcolor[HTML]{BDBDBD}Acronym & \cellcolor[HTML]{BDBDBD}Definition  \\ [0.5ex] 
 \hline
5G & fifth generation \\
\hline
5G-NR & 5G new radio \\
\hline
6G & sixth generation \\
\hline
AD & activity detection \\
\hline
AMP & approximate message passing \\
\hline
AOA & angle of arrival \\
\hline
AWGN & additive white Gaussian noise \\
\hline
BCH & Bose–Chaudhuri–Hocquenghem \\
\hline
BP & belief propagation \\
\hline
BPSK & binary phase shift keying \\
\hline
BS & base station \\
\hline
CB-MLD & covariance-based MLD \\
\hline
CCS & coded/coupled compressive sensing \\
\hline
CDMA & code division MA \\
\hline
CP & cyclic prefix \\
\hline
CRC & cyclic redundancy check \\
\hline
CS & compressed sensing \\
\hline
CSA & coded slotted ALOHA \\
\hline
CSI & channel state information \\
\hline
DFT & discrete Fourier transform \\
\hline
DOF & degree of freedom \\
\hline
FBL & finite blocklength \\
\hline
FDMA & frequency division MA \\
\hline
FEC & forward error correction \\
\hline
FFT & fast Fourier transform \\
\hline
FL & federated learning \\
\hline
GB-CR & graph-based channel reconstruction and collision resolution\\
\hline
GMAC & Gaussian MA channel \\
\hline
gOMP & generalized orthogonal matching pursuit \\
\hline
HWI & hardware impairments \\
\hline 
HyGAMP & hybrid generalized AMP\\
\hline
i.i.d. & independent and identically distributed \\
\hline
IDMA & interleave-division multiple access \\
\hline
IISD & iterative inter-symbol decoder \\
\hline
IoT & Internet-of-things \\
\hline
IRSA & irregular repetition slotted ALOHA \\
\hline
ISAC & integrated sensing and communication \\
\hline
LASSO & least absolute shrinkage and selection operator \\
\hline
LDPC & low-density parity check \\
\hline
LLR & log-likelihood ratio \\
\hline
LMMSE & linear minimum mean square error \\
\hline
LS & least squares \\
\hline

\end{tabular}
\hfill
\begin{tabular}[t]{|p{0.13\linewidth}|p{0.31\linewidth}|} 
 \hline
 \cellcolor[HTML]{BDBDBD}Acronym & \cellcolor[HTML]{BDBDBD}Definition  \\ [0.5ex] 
 \hline
MA & multiple access \\
\hline
MAC & MA channel \\
\hline
MIMO & multiple-input multiple-output \\
\hline
ML & machine learning \\
\hline
MLD & maximum likelihood detection \\
\hline
MMSE & minimum mean square error \\
\hline
mMTC & massive machine-type communications \\
\hline
MMV-AMP & multiple measurement vector AMP \\
\hline
MRC & maximal ratio combining \\
\hline
MSE & mean square error \\
\hline
MS-MRA & multi-stage setup with multiple receive antennas \\
\hline
MS-MRA-WOPBE & MS-MRA without pilot bits encoding \\
\hline
MSUG & multi-stage setup with user grouping \\
\hline
MSUG-MRA & MSUG for multiple receive antennas \\
\hline
MTC & machine-type communications \\
\hline
MUSIC & multiple signal classification \\
\hline
NNLS & non-negative least squares \\
\hline
NOMA & non-orthogonal MA \\
\hline
NP & Neyman-Pearson \\
\hline
ODMA & on-off division MA \\
\hline
OFDM & orthogonal frequency division multiplexing \\
\hline
OMA & orthogonal MA \\
\hline
OMP & orthogonal matching pursuit \\
\hline
PMF & probability mass function \\
\hline
PUPE & per-user probability of error \\
\hline
QPSK & quadrature phase shift keying \\
\hline
RA & repeat accumulate \\
\hline
RIS & reconfigurable intelligent surface \\
\hline
RS & Reed-Solomon \\
\hline
SCLD & successive cancellation list decoder \\
\hline
SCMA & sparse code MA \\
\hline
SDMA & space division MA \\
\hline
SIC & successive interference cancellation \\
\hline
SINR & signal-to-interference-plus-noise ratio \\
\hline
SISO & soft-input soft-output \\
\hline
SKP & sparse Kronecker product \\
\hline
SNR & signal-to-noise ratio \\
\hline
SPARC & sparse regression code \\
\hline 
TDMA & time division MA \\
\hline
TIN & treating interference as noise \\
\hline
UE & user equipment \\
\hline
UNISAC & unsourced ISAC \\
\hline
URA & unsourced RA \\
\hline
\end{tabular}
\label{table_acron}
\normalsize
\end{table*}

\subsection{\textcolor{black}{Structure of the Survey}}
The remainder of the survey is organized as follows. In Section II, a general discussion of MA techniques is provided. We begin with coordinated MA techniques and progress to random access. In Section III, we focus on URA as a new paradigm for uncoordinated MA communications with a massive number of potential devices, enumerate its advantages, and hint at some solutions developed in the literature to address the related challenges. Sections IV and V are dedicated to in-depth overviews of the existing URA solutions, with Section IV focusing on URA over GMAC and Section V on URA over fading channels. Regarding communication models over practical wireless channels, we first investigate the works that take into account channel fading with a single antenna receiver and describe the general approaches; then, we discuss those with a MIMO assumption. Finally, we conclude the survey in Section VI with remarks on URA solutions and provide future research directions. \textcolor{black}{The structure and organization of the paper are illustrated in Fig. \ref{structure}.} \textcolor{black}{Unlike prior surveys that focus primarily on GMAC, our structure integrates both single-antenna and MIMO fading URA solutions, thereby offering a comprehensive view that highlights practical deployment challenges.}
\textcolor{black}{While our survey spans the full URA literature, the unifying application context is 6G and IoT, where sporadic traffic and massive device density make URA particularly relevant.}

The acronyms used in this paper are alphabetically summarized in Table \ref{table_acron}.

\section{Background: Multiple Access}
Multiple access refers to the scenario in which two or more users simultaneously share the system resources. In this section, we provide a concise overview of the main MA approaches and outline their pros and cons to better understand the position of URA solutions in next-generation communication systems.

\subsection{Coordinated Access}
One way to serve multiple users is to assign dedicated resources (e.g., time, frequency, code, or their combinations) to them by a central coordination unit through a prior handshaking procedure. The conventional techniques in this framework include time-division MA (TDMA), frequency-division MA (FDMA), orthogonal frequency-division MA (OFDMA), code-division MA (CDMA), and space-division MA (SDMA). In TDMA and FDMA, non-overlapping time slots and frequency sub-channels are assigned to the different users, respectively, and each user utilizes its allocated time frame or frequency band for transmission \cite{Biglieri2007}. On the other hand, in CDMA, all users can simultaneously use all time-frequency resources through different (nearly) orthogonal code sequences, coupled with a low-complexity decorrelation-based receiver. OFDMA provides frequency division multiplexing by assigning different sets of subchannels to the individual users in an orthogonal frequency division multiplexing (OFDM) system, while SDMA relies on the principle of creating (nearly) orthogonal spatial channels for different users to minimize the inter-user interference by employing beamforming and spatial multiplexing \cite{Shah2021Survey}.

All the approaches described in the previous paragraph fall under the category of OMA, where the number of users is strictly limited due to orthogonal resource assignment. To increase the number of supported users, non-orthogonal multiple access (NOMA) techniques have also been investigated in the literature, where multiple users can utilize the same resources to transmit their data, thereby allowing the system to support a greater number of users. However, this imposes inter-user interference on the system, and more sophisticated interference-canceling receivers are needed with an increased complexity compared to OMA.

NOMA schemes are divided into two main categories: power-domain NOMA and code-domain NOMA. In power-domain NOMA \cite{pdnoma}, the users employ different power levels to transmit their data using the same resources, and the receiver exploits the received power differences for detection through SIC. On the other hand, code-domain NOMA is similar to classical CDMA in that users employ user-specific code sequences to transmit their data. The code sequences are sparse or non-orthogonal sequences with low cross-correlations \cite{surveynoma}. Here, the receiver can employ message-passing algorithms that utilize the code structure to differentiate between users. Some examples of code-domain NOMA schemes in the literature are low-density spreading code division MA \cite{cdnoma}, low-density spreading aided orthogonal frequency division multiplexing \cite{ldsofdm}, sparse code MA (SCMA) \cite{scma}, and multiuser shared access \cite{musa}. 

\subsection{Random Access}

All of the schemes described in the previous subsection are coordinated. That is, they require scheduling grants and coordination with the base station (BS) before transmission, and it is assumed that the number of users and their activity patterns are available at the BS. However, when the number of users becomes large, they become infeasible due to the excessive delays and signaling overhead. For example, the typical number of users per signal dimension in the time-frequency domain (per degree of freedom) is between 1.5 and 3 according to 3GPP studies, as the system performance significantly degrades beyond this threshold \cite{survey}. The common solution addressing these drawbacks is random access, where the users perform their transmission without any coordination by the BS whenever they have data. ALOHA and grant-free random access are the two common RA approaches in the literature.

\subsubsection{\textbf{ALOHA \& Variants}}
 
\begin{figure}[t]
    \centering
    \includegraphics[trim={0cm 6cm 7cm 1.5cm},clip,width=1\linewidth]{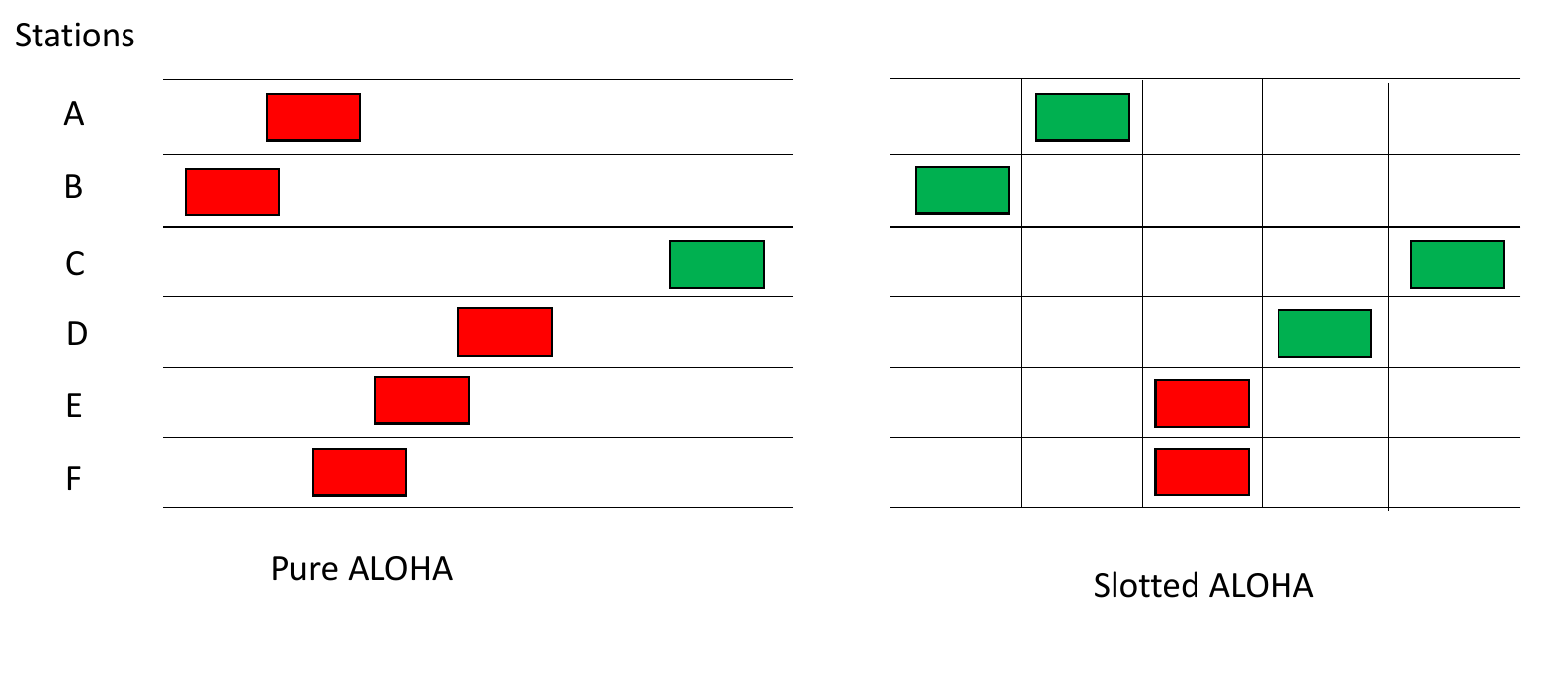}
    \caption{An illustration of pure and slotted ALOHA. Red boxes show the collided packets, and green boxes are the successfully transmitted ones.}
    \label{figaloha}
\end{figure}

An early RA approach is the ALOHA protocol \cite{Abramson1970The} (also known as \textit{pure ALOHA}) proposed in the 1970s, 
where a device sends the entire packet at a randomly selected time. If no acknowledgment is received within a certain period of time, it is assumed that a collision has occurred with a packet sent by another device, and the packet is retransmitted after an additional random waiting time to avoid repeated collisions. This process is repeated until a successful transmission and its subsequent confirmation occur or until the process is terminated by the user console. Slotted ALOHA (SA) is a variant of ALOHA, where packets are transmitted in designated time slots, thereby doubling the maximum throughput compared to pure ALOHA \cite{Roberts1975ALOHA}. 
 However, collision is still a significant challenge in SA, as several active users may share the same slot to send their data packets, leading to packets colliding within that slot and the discarding of data. The transmission structures in pure ALOHA and slotted ALOHA are illustrated in Fig. \ref{figaloha}. 

With the motivation to increase the throughput, several variants of SA have been proposed in recent literature, where the main idea is to transmit multiple copies of a packet in different time slots. If the packet is successfully recovered in one slot, the effect of its replicas in the other slots is removed. Based on this idea, two copies of the same packet are transmitted in contention resolution diversity slotted ALOHA (CRDSA) \cite{Casini2007Contention}, and the packet repetitions are picked from a probability mass function optimized for higher throughput in irregular repetition slotted ALOHA (IRSA), proposed in \cite{Liva2011Graph}. An extension of IRSA, exploiting forward error correction, is called coded slotted ALOHA (CSA) \cite{Paolini2015Coded}. This approach achieves a throughput of almost one when sufficiently low coding rates are employed. IRSA and CSA are further studied in \cite{Choudhury1983Diversity, Irregular2019Irregular,Akyildiz2021Energy,Haghighat2023Analysis,Haghighat2023AnEnerg}.
Moreover, the resolution of partial collisions in random access setups is discussed in \cite{Kazemi2022Collision}.


\subsubsection{\textbf{Grant-Free Random Access}}
Grant-free RA is another RA candidate that enables communication among a large number of users, where users transmit their data by utilizing shared system resources without requiring a prior grant from the BS, thereby reducing latency and signaling overhead. Nevertheless, the users can have unique signatures/preambles or they can utilize different codebooks, which means that there is user identity during the data transmission phase, and the receiver can use this information for activity detection or user separation \cite{Pyo2017Athrough}. This is also called \textit{sourced random access}. Among the grant-free RA solutions, grant-free NOMA is a prominent solution for future communication systems due to its low latency, improved spectral efficiency, and potential to support massive connectivity. Note that the main difference between grant-based and grant-free NOMA is that the BS does not have information on the set of active users in the latter, which brings significant challenges requiring different solutions such as blind activity detection, channel estimation, data recovery, and synchronization with minimal overhead \cite{shahab2017grant}.

Among the two main categories of NOMA, power-domain NOMA is challenging to extend to the grant-free scenario, as the successful recovery of transmitted packets at the receiver heavily depends on the power difference between users, which may not be maintained due to the grant-free nature of the transmissions. On the other hand, signature-based schemes in code-domain NOMA can be adapted to the grant-free case with proper activity detection before the data detection or joint activity detection and data recovery \cite{Razavi2016Non}. For instance, a grant-free version of SCMA is proposed in \cite{scmagf}. Furthermore, the inherent sparsity of the user activity in grant-free MA is utilized in \cite{csgf1,csgf2,csgf3,liu} through compressive sensing algorithms for multiuser detection.

\section{URA: A Recent Paradigm}

\begin{figure}[t]
    \centering
       \includegraphics[trim={0 -0.5cm 0 0},clip,width=0.8\linewidth]{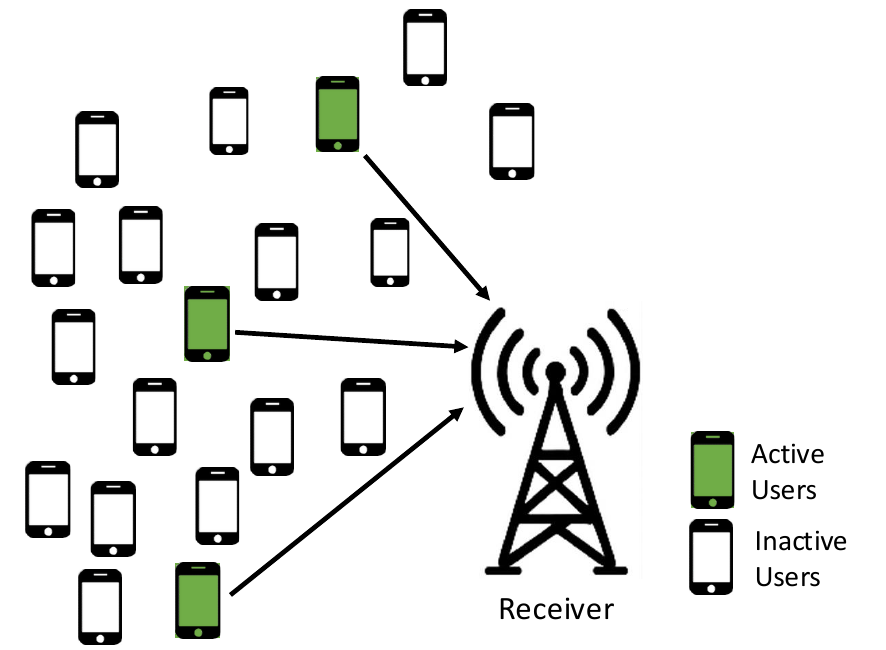}
    \caption{An illustration of a system where a large number of potential users (devices) try to communicate with a receiver (access point), with only a small subset of them being active at each time frame.}
    \label{fig_sporadic}
\end{figure}

In some next-generation communications applications, such as massive MTC and massive IoT, the total number of devices can be in the order of millions, and they are only active very sporadically, as illustrated in Fig. \ref{fig_sporadic}. This makes achieving any level of coordination infeasible. As a result, none of the MA techniques summarized in the previous section is suitable for enabling the communication of these devices. Specifically, the complexity of the grant-free MA is prohibitive due to the codebook size when each potential user is assigned a unique preamble. On the other hand, ALOHA leads to excessive collisions, significantly degrading communication efficiency. 

To address this problem, Polyanskiy proposed the URA paradigm in \cite{polyanskiy2017perspective} where the data transmission and user identification problems are decoupled via the utilization of a common codebook by all users. Therefore, the receiver is not interested in user identities during data transmission, with decoding being performed only up to a permutation of the transmitted messages.

\subsection{General System Model}
\textcolor{black}{For brevity, we present the general system model for a URA system with a multiple-antenna receiver and complex channel coefficient, since conversions to real channel coefficients and a single-antenna receiver can easily be obtained as special cases. We also assume a quasi-static channel, where the channel coefficients remain constant over the transmission frame \footnote{\textcolor{black}{The transmission frame is the collection of available resources in different dimensions, such as time and frequency, to be used for transmission. We call it \textit{frame} for brevity hereafter.}}, as the setup can easily be extended to cases where the channel coefficients vary over the frame by replacing the frame with a coherence time block or a time slot over which the channel remains constant. Note that these assumptions are made for the sake of brevity in presenting the general URA system model; however, we have investigated all existing URA solutions in the literature, regardless of their specific system models.} Considering a URA system with an $M_r$-antenna receiver and single-antenna users, in which $K_a$ out of $K_T$ users are active at a given frame of length $n$ \cite{polyanskiy2017perspective}, the received signal, $\mathbf{Y} \in \mathbb{C}^{M_r\times n}$, in the absence of synchronization errors can be written in the following general form \cite{Ahmadi2023Unsourced}
\begin{align}
	\mathbf{Y} = \sum_{i=1}^{K_a} \mathbf{h}_i{\mathbf{x}_i(\mathbf{u}_i)+{\mathbf{Z}}}\in \mathbb{C}^{M_r\times n},
	\label{eqs1_Sec2}
\end{align}
where $\mathbf{x}_i(\mathbf{u}_i)$ is the $1 \times n$ transmitted signal of the $i$-th user corresponding to its message bit sequence $\mathbf{u}_i\in \{0,1\}^B$ of length $B$, ${\mathbf{Z}}\in \mathbb{C}^{M_r\times n}$ is the circularly-symmetric additive white Gaussian noise (AWGN) with independent and identically distributed (i.i.d.) elements drawn from $ \mathcal{CN}({0}, \sigma_z^2)$, with $\sigma_z^2$ being the noise power, \textcolor{black}{ and ${\mathbf{h}}_i\in \mathbb{C}^{M_r\times 1}$ is the vector of channel coefficients. Note that in the URA literature over GMAC, a single-antenna receiver and unit channel coefficients are assumed. However, the default assumption for fading channels is quasi-static fading; i.e., the channel coefficients remain constant across a frame and change from one frame to the next.}

In the URA literature, it is assumed that each user selects its message index uniformly from the set $\{1,2,...,2^B\}$. The objective in URA is to determine the list of transmitted messages, $\mathbf{u}_i$'s, $ \forall i=1,2,...,K_a$, by processing the received signal $\mathbf{Y}$ so that the required energy-per-bit for a target per-user probability of error (PUPE), as the performance metric, is minimized \cite{vem2019user,Gkagkos2023FASURA,fading4}.
\textcolor{black}{The PUPE captures both missed detections and false alarms. Specifically, it is defined as the sum of probabilities that an active user’s message is not correctly included in the decoded list (probability of misdetection, $p_{md}$), and that an invalid message appears in the list (probability of false-alarm, $p_{fa}$);} i.e., \cite{Ozates2023Aslotted}
\begin{align}
	P_e = p_{fa}+p_{md},
\end{align}
\noindent where $p_{md} =\dfrac{1}{K_a}\sum_{i\in\mathcal{K}_a}^{}{ \mathrm{Pr}(\mathbf{u}_i\notin \mathcal{L}_d)}$, where $\mathcal{K}_a$ and $\mathcal{L}_d$ are the set of active user indices and the list of decoded messages, respectively, and $  p_{fa} = \mathbb{E}\left\{\dfrac{n_{fa}}{|\mathcal{L}_d|}\right\}$, with $n_{fa}$ being the number of decoded messages that were indeed not sent.

The energy per bit can be written as 
\begin{equation}
    \dfrac{E_b}{N_0}=\dfrac{nP}{c\sigma_z^2 B},
\end{equation}
where $P$ denotes the average power of each user per channel use, i.e., $\norm{\mathbf{x}_i}^2 \leq nP$, and we have $c=2$ for real-valued and $c=1$ for complex-valued channels.

\subsection{Challenges}
In the following sections, we review existing URA solutions in the literature. However, to gain a better understanding of the URA setup and the reasoning behind the solutions provided in the literature, we first need to identify the main challenges faced in a URA setup, which will be discussed in the remainder of this section.

\textbf{Interference:} 
Interference stands out as the primary challenge in MA channels, especially in random access setups such as URA, due to the lack of coordination between the receiver and the users. 
Failure to incorporate strategies to reduce or cancel it can lead to significant performance degradation. To demonstrate the detrimental impact of interference, we analyze the treating interference as noise (TIN) scheme, where single-user decoding is carried out by taking interference from other users as noise, without employing any special technique to mitigate it. The signal-to-interference-plus-noise ratio (SINR) of the $i$-th user, $\alpha_i$, in the TIN strategy for a GMAC scenario can be written as
\begin{align}
		\alpha_i = \dfrac{P}{\sigma_z^2+{(K_a-1)}P},
\end{align}
where $P$ is the transmit power of each user. Note that the SINR in a fading scenario is approximately the same as in \eqref{eqs1_Sec2} since we can approximate $\frac{1}{M_r}\|\mathbf{h}_i\|^2\approx 1, \ \forall i$, where $\|\cdot\|$ denotes the Euclidean norm. For a sufficiently large active user load $K_a$, the SINR can be written as
\begin{align}
		\alpha_i \approx \dfrac{1}{K_a}.
\end{align}

This equation illustrates that when the number of users exceeds a specific threshold (more specifically, when it is much greater than the inverse of the signal-to-noise ratio (SNR), i.e. $K\gg \frac{\sigma^2_z}{P}$, which is not a very large value in many envisioned URA applications), the SINR becomes inversely proportional to $K_a$. Therefore, boosting the transmit power does not result in an enhanced SINR, or equivalently, improved system performance. Consequently, in systems such as URA, the relatively large number of active users has an adverse impact on system performance. 

There are two main techniques in the URA literature to alleviate the effects of interference, with the most common being SIC, which removes the effects of successfully detected messages from the received signal to decrease the interference before another round of decoding. Especially in URA, SIC is a crucial step in achieving low error rates due to heavy multiuser interference. Other techniques for interference management involve randomly separating users into different groups, such as slotting the time frame, employing different transmit power levels, and multiplying or concatenating a known sequence to the data segment.
   \begin{enumerate}
       \item Time-domain separation: Dividing the frame into multiple slots and letting each user randomly select a slot to transmit its data can lead to improved system performance. In a slotted structure, the level of interference is reduced due to the lower number of interfering users in each slot. At the same time, the gain of channel coding is also decreased because fewer channel uses are available to each user. Thus, optimizing the number of slots based on system parameters, such as active user load, is crucial for maximizing the advantages of slotting.
       
       \item Code-domain separation: In some URA schemes, each user randomly selects a preamble/pilot/pattern and uses it to prepare its data for transmission. One way is to transmit a concatenation of a preamble or pilot and the main signal.
       At the receiver end, the channel coefficients for different users can be estimated using the detected pilots. Then, minimum mean square error (MMSE) or maximum ratio combining (MRC) solutions can be employed to separate the user packets, before decoding them using single-user channel decoders. 
       Another way is to multiply a randomly selected pattern by the main signal, which is used in random-spreading-based solutions.
       However, employing these techniques usually decreases the amount of power and time resources dedicated to the data (main signal) part. Therefore, for interference management to be effective, the choices of type, length, and power of the pilots and pattern sequences are crucial.

       \item Power-domain separation: An alternative approach for user separation, especially effective in GMAC scenarios, is the so-called \textit{power diversity} technique \cite{ahmadi2021random}, where each user randomly picks a transmit power level from a limited set with a specific probability distribution, satisfying the overall power constraint. Since the users are now separated by different power levels, the receiver, with a high probability, decodes the users with the highest power level first, and after each SIC step, it moves to decode the lower power levels. The appropriate design of the power diversity parameters, namely, the number of power levels, their values, and their probability distribution, is crucial for optimizing the system performance.

   \end{enumerate}
\textbf{Collisions:} 
In many classical random access setups, such as ALOHA, collision is used interchangeably with interference, i.e., when user packets are fully or partially received at the same time at the receiver. However, in the URA literature, collision by default refers to the collision of either user messages or user pilots, which occurs when two or more users have the same messages or pilot sequences, respectively. 
\begin{enumerate}
    \item Message collision: In the case of user message collision, with the help of a randomized encoding process independent of user messages, the receiver can recover collided user messages to some extent. Note that in a conventional URA setup, since the receiver is only interested in the list of transmitted messages, being able to recover at least one of the collided messages does not increase the misdetection rate. However, this matter becomes more important in some URA systems where the receiver is also interested in the frequency of user messages.
    \item Pilot collision: Many existing works taking into account channel fading employ preambles/pilots to estimate the channel coefficients used in the decoding phase. However, it is impossible to assign dedicated preambles/pilots due to the massive number of potential users. Therefore, as a common approach, the users independently pick a preamble/pilot from a common codebook, either randomly or based on a part of their message bits. Hence, if the pilot bits of multiple users are the same (called pilot contamination in massive MIMO literature \cite{bjornson2017massive}), a collision occurs, and the system struggles to decode the transmitted packets. Therefore, it is crucial to perform a preliminary analysis to optimize the design parameters to keep the pilot collision probability lower than a desired threshold or develop techniques for its mitigation.
\end{enumerate}

\textbf{Computational Complexity:} 
In order to increase the number of active users supported by a URA system, larger values of parameters such as frame length and the number of antenna elements (in MIMO systems) are necessary. However, the computational complexity of URA solutions typically increases superlinearly with these parameters, as well as the number of active users.
As a result, solutions for a URA system with high dimensions, including temporal (frame length), spatial (number of receiver antennas), and/or the number of active users, are expected to experience high computational complexity. To mitigate this complexity, various strategies are employed in the literature, such as slotted transmissions and coded compressed sensing (CCS). The idea of CCS is to divide the bit sequence into smaller segments, map each segment to a column of a common codebook, and stitch them together using a tree code, rather than mapping the entire bit sequence to the columns of a $2^B \times n$ codebook, which would lead to prohibitive receiver complexity. This reduction in dimensionality helps lower the computational load at the decoder, albeit at the cost of reduced performance.

{\color{black} In this section, we summarized the main challenges in a URA setup along with a high-level discussion on the proposed solutions in the literature. In the two subsequent sections, we provide a more detailed examination of these solutions, with Section IV dedicated to GMAC and Section V covering single-antenna (V.A) and MIMO (V.B) fading setups.
We take a systematic approach in presenting the URA solutions in these sections: different contributions are grouped based on their solution approach. In each group, the underlying approach is explained with an emphasis on the main structure of its encoding and decoding approaches, and the relevant works are presented in chronological order. To provide an overall view, at the end of each main subsection, a summary of the primary contributions is presented, and a comparative discussion of their performance is given, accompanied by some numerical results.}

\section{URA over GMAC}
In this section, we provide an overview of URA works over GMAC. To this end, we first describe an achievability bound on the performance of the URA over GMAC, as developed in Polyanskiy's original paper \cite{polyanskiy2017perspective}, and then review the existing URA over GMAC solutions to achieve it. 

\subsection{Performance Bounds}

\textcolor{black}{Polyanskiy provides a random coding achievability bound on the performance of URA over GMAC in \cite{polyanskiy2017perspective}, as a combination of a Chernoff-type bound and an information density-based one, which has been used as a benchmark for the energy efficiency in the URA over GMAC literature. }
The random coding bound is an information-theoretic performance limit, derived without any complexity constraints. It assumes that all active users are jointly decoded, which has a prohibitive complexity. So, practical coding schemes for URA with reasonable complexity are needed to achieve the results predicted by the bound. 




 In \cite{polyanskiy2017perspective}, the performances of common RA solutions are evaluated in a URA setup, assuming that there are $K_a$ active users transmitting $B = 100$ bits of information through a transmission frame of length $n = 30000$, with a target PUPE of $\epsilon = 0.1$. Note that the same set of parameters has been adopted by subsequent URA works over GMAC for performance comparison, for obvious reasons. However, a more comprehensive analysis of the existing competing solutions, especially those nominated for inclusion in future protocols, is needed that covers different sets of parameters representing different future URA applications.
 
\begin{figure}
    \centering
   \includegraphics[width=1\linewidth]{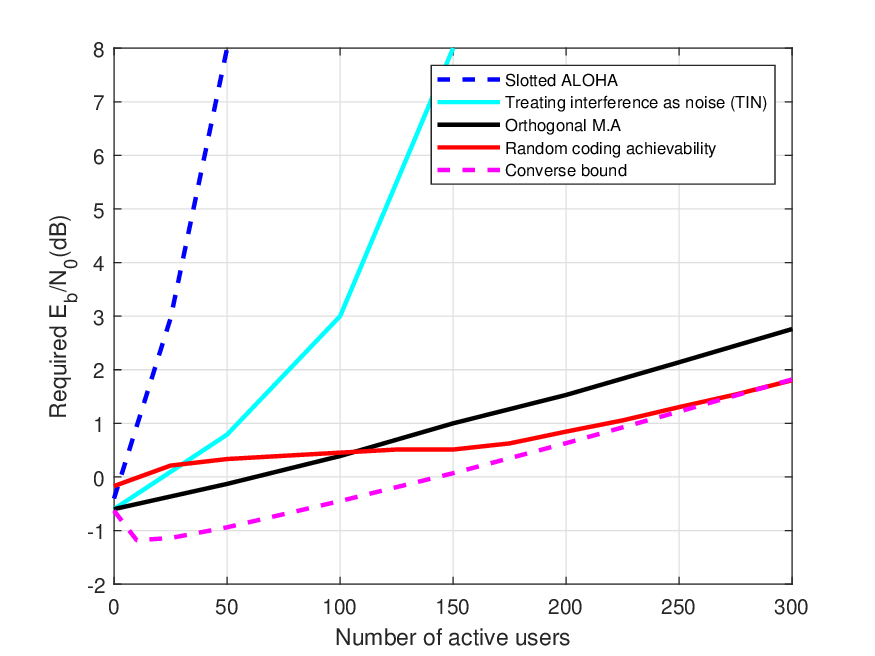}
        \caption{Performance comparison of conventional random access techniques with Polyanskiy's URA (random coding) achievability bound.}
    \label{fig_eff_poly}
\end{figure}

 The curves for the required $E_b/N_0$ versus the number of active users for slotted ALOHA, TIN, and orthogonal MA are plotted in Fig. \ref{fig_eff_poly} along with the random coding bound derived in \cite{polyanskiy2017perspective} and a converse bound for $K_a$-GMAC from \cite{MolavianJazi2015ASecond}. The results in Fig. \ref{fig_eff_poly} show that the conventional RA techniques (i.e., slotted ALOHA and TIN) are energy inefficient in a URA setup, as there is a huge performance gap between these and the random coding bound, and the development of low-complexity coding schemes is needed. 
\textcolor{black}{For slotted ALOHA, it is assumed that a user packet is successfully recovered if there is no collision and the single-user decoding is successful, while the finite blocklength capacity of TIN is estimated using the normal approximation in \cite{Polyanskiy2010Channel}.}
 The performance of the grant-based TDMA scheme is also plotted in the figure (referred to as \textit{Orthogonal M.A}) as a benchmark, in which the frame is divided into sub-frames of length $\frac{n}{K_a}$ and each active user transmits in its dedicated sub-frame, with its required $E_b/N_0$ being calculated using the normal approximation in \cite{Polyanskiy2010Channel}. However, note that the TDMA scheme is not feasible in a URA setup, mainly due to the lack of coordination. Leaving the coordination issue aside, the system works in the regime of $n<K_a B$ in some URA applications, rendering orthogonal solutions such as TDMA infeasible.

The random coding bound in \cite{polyanskiy2017perspective} is developed under the assumption that the receiver knows the number of active users $K_a$. The authors in \cite{ngo2023unsourced} extend this bound to the case where the number of active users is random and unknown at the receiver. Furthermore, an achievability bound for the binary-input signaling for URA over GMAC is derived in \cite{glebov2023energy}. A random coding bound for the scenario that the users send a common alarm message in addition to standard messages is developed in \cite{ngo2024unsourced}, where the alarm message needs to be decoded with a lower error rate than the standard ones. An achievability bound for the case of asynchronous GMAC is obtained in \cite{Wu2024}, where the delay in transmission is assumed to be up to a small portion of the time frame. However, the provided achievability bound is loose for the asynchronous case as the authors have employed a fixed frame decoder in which all the packets that have some overlaps with the end of the frame are deemed undecodable, leading to huge performance degradation, which is not valid in a fully asynchronous setup, where the time horizon is unbounded. \textcolor{black}{The drawbacks in \cite{Wu2024} are alleviated in \cite{Jyun2025Wrap}, where the authors study a wrap-around version of the model in \cite{Wu2024}, that is, the last $D_m$ symbols are added to the first $D_m$ symbols, where $D_m$ is the maximum delay. As a result, the receiver observes the superposition of $K_a$ cyclically shifted codewords with Gaussian noise, where the noise power in the first $D_m$ symbols is doubled. In this way, the energy efficiency of the bound in \cite{Wu2024} is significantly improved.} \textcolor{black}{The problem of joint delay and activity detection in the asynchronous setup is investigated in \cite{Derya2024} using methods of tropical geometry, which is the study of the geometric properties of polynomials when addition is replaced by minimization and multiplication by ordinary addition. Moreover, the conditions for path delays for zero-error detection are identified.} \textcolor{black}{Furthermore, the authors in \cite{Ritter2025Removal} examine a two-user binary adder channel in the asynchronous scenario, and propose an algorithm to avoid the small weight stopping sets in the joint factor graph, which mitigates the error floor due to these sets.}

\subsection{URA Solutions for GMAC}
As shown in the previous subsection, the conventional RA techniques are energy-inefficient for the URA framework. As a result, there have been substantial efforts to develop low-complexity and energy-efficient schemes that can be used in practical systems. \textcolor{black}{The main approaches for this purpose in the literature can be categorized as slotting the transmission frame, coded compressed sensing (CCS), random spreading, interleave-division MA (IDMA), and on-off division MA (ODMA). In the following, we review these schemes.} 

\subsubsection{\textbf{\textcolor{black}{Slotting-based schemes}}}
One of the design strategies is slotting the transmission frame, where each active user randomly picks one or a few slots to transmit its data, depending on the exact setting. In this direction, Ordentlich and Polyanskiy propose the first practical scheme for URA in \cite{ordentlich}, where a concatenated coding scheme combined with the $T$-fold ALOHA protocol is considered. $T$-fold ALOHA is an approach similar to slotted ALOHA in the sense that the users choose a random sub-block (slot) to transmit their packets, but in slotted ALOHA, packets are lost in the case of a collision. However, in $T$-fold ALOHA, $T$ or fewer users can be simultaneously decoded in each slot. In \cite{ordentlich}, each user randomly chooses one slot to transmit its message. This concatenated coding scheme consists of an inner binary linear code, which is used to decode the modulo-2 sum of the codewords transmitted within a slot, and an outer code to recover the individual messages. Many off-the-shelf codes can be used for the inner code. The authors construct the outer code from the columns of a $T$-error correcting Bose–Chaudhuri–Hocquenghem (BCH) code. This scheme suffers in terms of energy efficiency as the performance gap with the random coding bound is about 20 dB when the number of active users is 300. Its performance is significantly improved in \cite{zhang2024improved} (\textcolor{black}{by up to 18 dB}), where the authors employ a concatenated coding scheme with polar and BCH codes as inner and outer codes, respectively, and propose a decoding approach that exchanges information between inner and outer decoders. \textcolor{black}{Specifically, all the candidate codewords from the inner successive cancellation list decoding of the polar code, rather than just one, are passed to the outer BCH decoder to facilitate its error checking. In addition, indicators for the decoding status of the outer BCH decoder are used to enhance the inner polar decoding process.}

\begin{figure}
    \centering
    \includegraphics[trim={0cm 9cm 11.5cm 0},clip,width=1\linewidth]{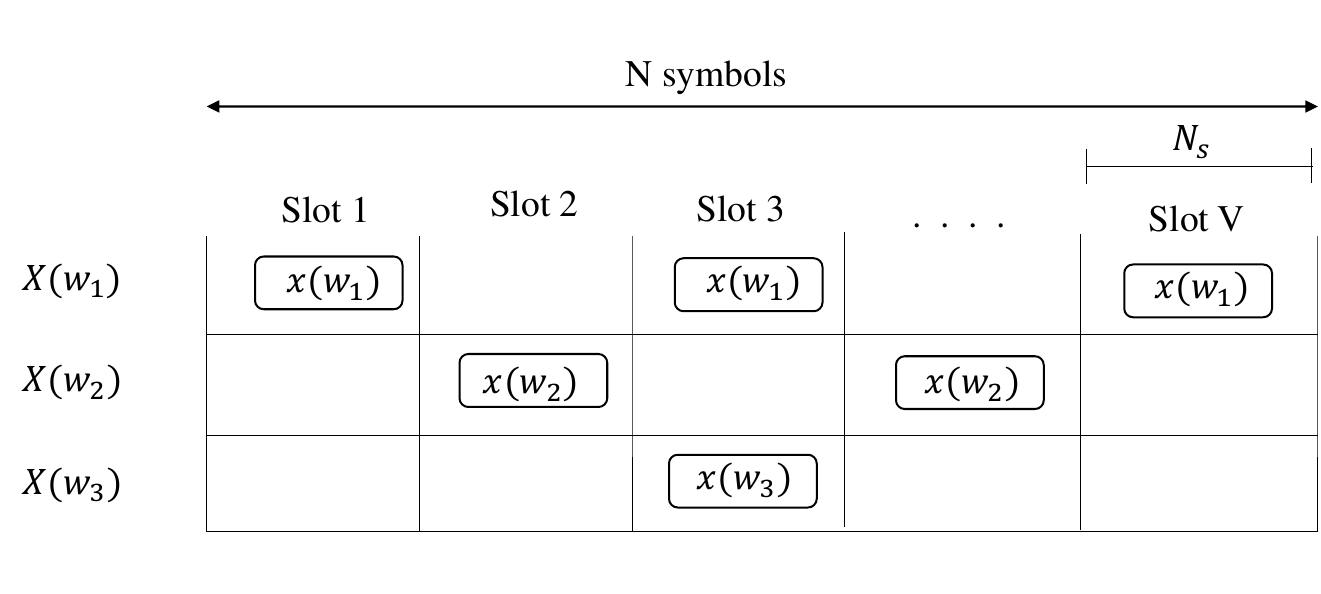}
    \caption{The slotted transmission structure in \cite{vem2019user} based on $T$-fold IRSA.}
    \label{figslot}
\end{figure}

The authors in \cite{vem2019user} consider a slotted structure where the users can repeat their codewords across sub-blocks, which is referred to as \textcolor{black}{$T$}-fold IRSA. They propose splitting the message into two parts and employing a coding scheme based on a combination of compressed sensing (CS) \cite{clazzer2024investigation} and LDPC coding at the slot level. Namely, the users select a preamble determined by the first part of the message from a common codebook, and the second part is encoded using an LDPC code and interleaved before transmission. The preamble carries information about the interleaver and the utilized slots for a specific user. At the receiver side, the preamble part is recovered by a non-negative least squares (NNLS) algorithm, and the coded part is decoded by joint belief propagation (BP). The replicas of the recovered codewords are peeled from the other slots via SIC. \textcolor{black}{This scheme improves the performance of \cite{ordentlich} by up to 13 dB, and the difference with the random coding bound is less than 8 dB.} The general transmission structure in $T$-fold IRSA based URA solutions is depicted in Fig. \ref{figslot}, \textcolor{black}{where $N$ is the length of the transmission frame, $N_s$ is the slot length, $V$ is the number of slots, $w_i$ is the message of the $i$-th user, and $x(w_i)$ denotes the transmitted signal of the $i$-th user.}

$T$-fold IRSA is also utilized in \cite{marshakov2019polar} by replacing the LDPC codes with polar codes and utilizing joint successive cancellation decoding \cite{arikan2009Channel}, which provides a significant performance gain over \cite{vem2019user} \textcolor{black}{due to the superiority of the polar codes to the LDPC codes in the short blocklength regime.} In \cite{Tanc20Massive}, the authors employ random signatures to convert the equal-gain channel into a more favorable i.i.d. fading scenario along with trellis-based codes in a $T$-fold ALOHA structure. Their proposed approach provides a low-complexity solution with moderate performance. Moreover, the authors in \cite{sui2023finite} employ finite-field coding in a slotted structure. More specifically, the users encode their data using a non-binary LDPC code and transmit it after repetition, zero-padding, and interleaving. At the receiver, intra-slot decoding is performed using a joint belief propagation (BP) algorithm over a non-binary factor graph. Their proposed approach improves the performance of the \textcolor{black}{previously proposed} slotted \textcolor{black}{URA} schemes for high active user loads. 


    \subsubsection{\textbf{\textcolor{black}{CCS-based schemes}}}
\begin{figure}
    \centering
    \includegraphics[trim={4cm 6cm 4cm 1.5cm},clip,width=1\linewidth]{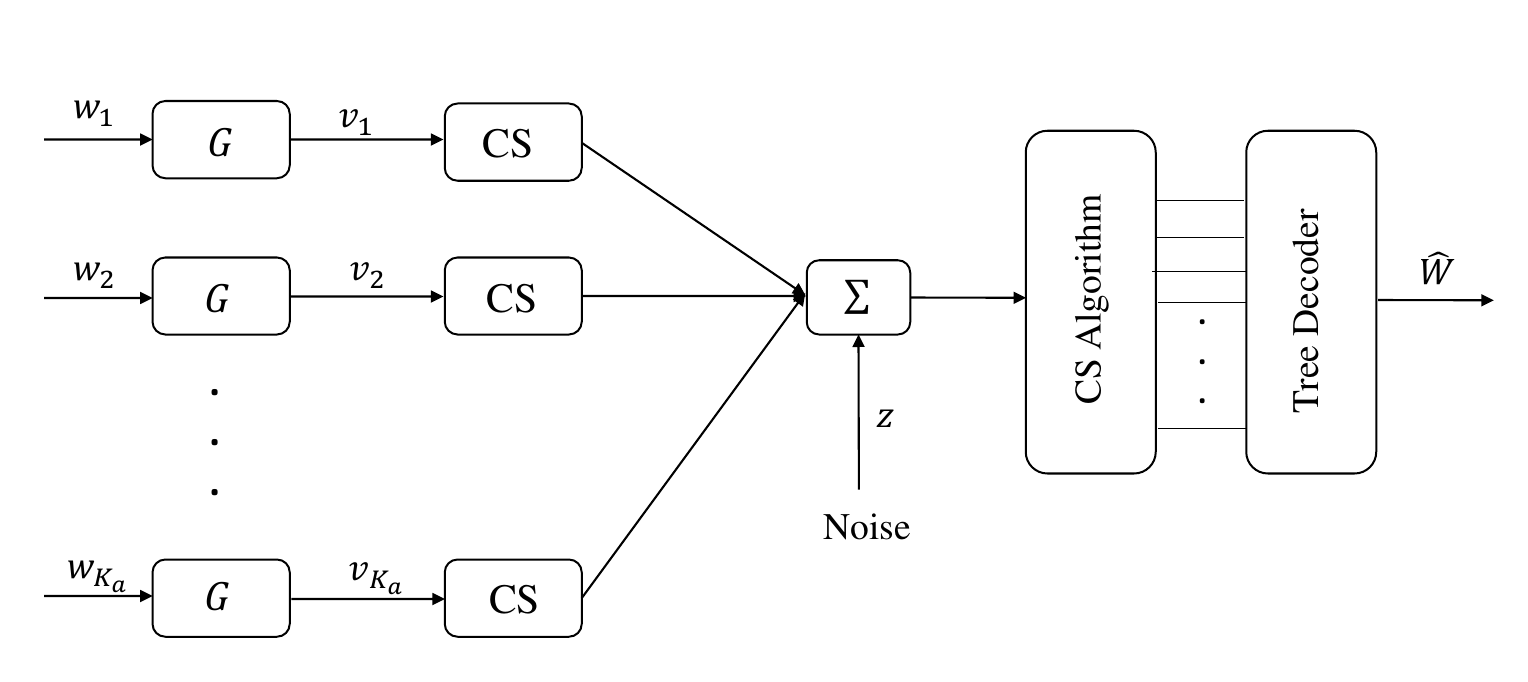}
    \caption{Encoding-decoding structure in a coded compressed sensing (CCS) approach.}
    \label{figccs}
\end{figure} 

\begin{figure*}
    \centering
    \includegraphics[trim={0cm 8.5cm 0cm 0},clip,width=0.9\textwidth]{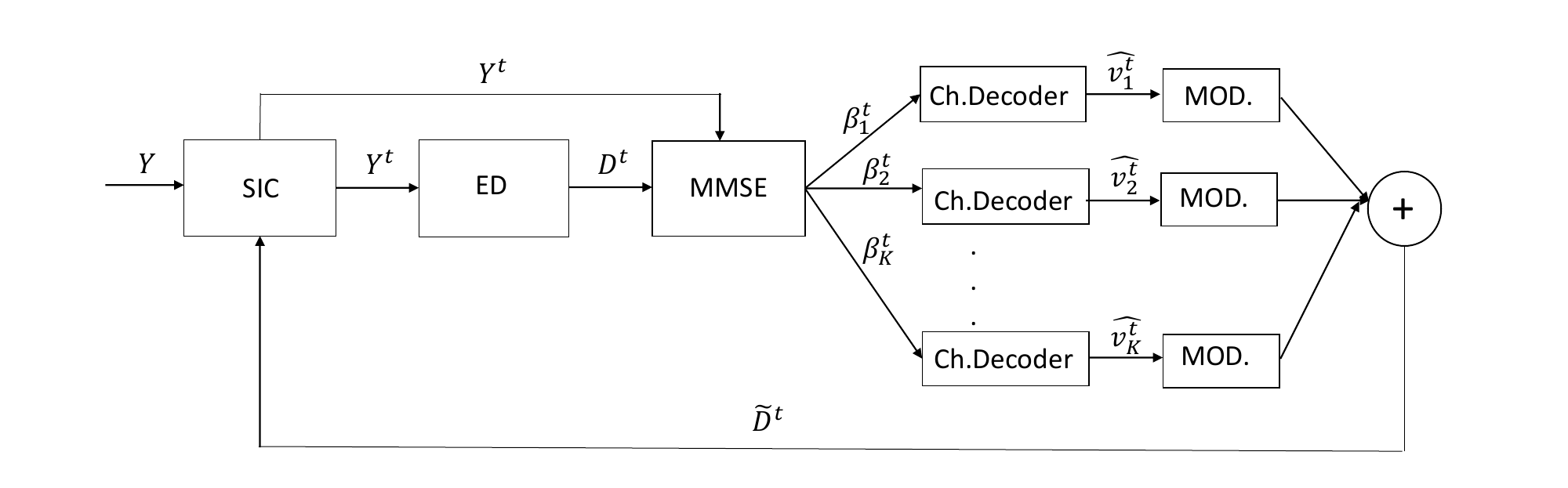}
    \caption{A typical decoding structure in a random spreading approach.}
    \label{figdecspread}
\end{figure*} 

Another main approach in URA over GMAC is called coded compressed sensing (CCS) \cite{Amalladinne2020Acoded,comp3,comp4,comp5,Agostini2023Constant,comp6,nassaji2024dynamic}, where the messages are divided into sub-blocks and each sub-block is first encoded by an outer code by appending parity bits before being mapped to a CS codebook for inner encoding. The encoded sub-blocks are then transmitted in different slots. At the receiver side, each sub-block is recovered via compressed sensing algorithms, and the recovered segments are stitched together by an outer code, \textcolor{black}{where a tree code is typically used}. The general encoding-decoding structure in coded compressed sensing is illustrated in Fig. \ref{figccs}, \textcolor{black}{where $G$ is a random binary matrix performing tree encoding, $v_i$ is the output signal of the tree encoder for the $i$-th user, and $\hat{W}$ is the set of decoded messages.} {\color{black} More specifically, the encoding process is performed in two phases: 1) In the tree encoder, $B$ bits of information are divided into $S$ parts, and all of these segments, except the initial one, are appended additional parity-check bits; 2) in the CS encoder, each $S$ sub-message is mapped to a CS codebook to select a sequence for transmission. Hence, the received signal is comprised of $S$ slots, with each slot being a superposition of the sequences from $K_a$ users. \textcolor{black}{Note that no parity-check bits are appended to the initial segment of the tree encoder since this segment is used as a root node in the tree decoding process \cite{Amalladinne2020Acoded}.} The decoder also functions in two stages: 1) In the initial stage, $K_a$ sequences associated with the $K_a$ active users are identified within each slot. 2) In the second stage, called outer decoder or tree decoder, with the assistance of the unique parity bits, the sub-messages corresponding to the detected sequences are combined to create a message of length $B$.}

\begin{figure*}
\includegraphics[width=1.\linewidth]{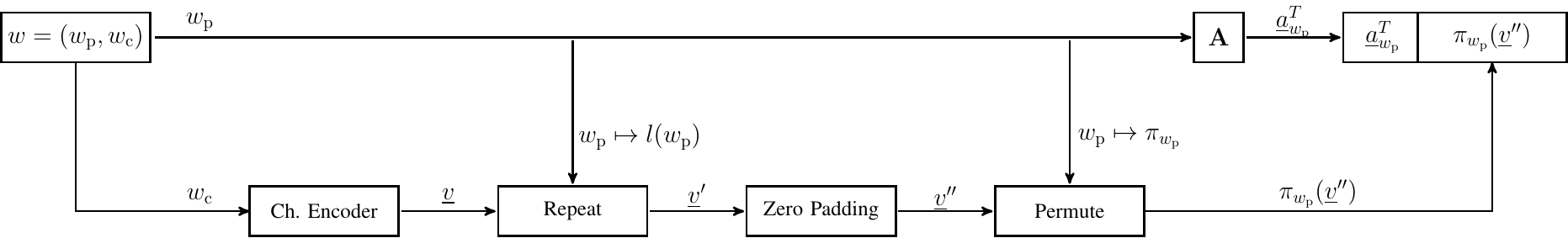}
\vspace{1mm}
		\caption{ Encoding process of sparse IDMA \cite{pradhan2022Sparse}.}
        \label{sparseIDMA}
\end{figure*}


Several variations of the CCS approach are examined in the URA literature. Specifically, in \cite{Amalladinne2020Acoded}, the sub-blocks are first encoded by \textcolor{black}{an outer tree code} and then mapped to the columns of a common sensing matrix for inner encoding. \textcolor{black}{In \cite{comp3}, sparse regression codes (SPARCs), a class of channel codes that can achieve channel capacity of a point-to-point AWGN channel under approximate message passing (AMP) decoding, are exploited as inner codes to improve the performance of the scheme in \cite{Amalladinne2020Acoded}, while the outer code is again a tree code}. The authors in \cite{comp4} further improve the performance of the approach in \cite{comp3} by introducing a dynamic interaction between the inner AMP and the outer tree decoder. SPARCs are also employed in \cite{comp5}, where the sub-blocks are CRC-encoded and connected via block Markov superposition transmission. At the receiver, a hybrid decoder combining SIC and AMP is used for inner decoding, with the addition of tree decoding. Furthermore, the authors in \cite{Agostini2023Constant} employ constant-weight codes as outer codes with the use of Gabor frames to design the inner CS coding matrix, and a concatenation of Raptor codes and tree codes is used as the outer code in \cite{comp6}. Finally, in a recent work \cite{nassaji2024dynamic}, the authors propose a CCS-based scheme eliminating the tree code by using the encoded sub-message in the first slot as a user identifier. This reduces the parity bits and the resulting rate loss, offering the best performance among the CCS-based schemes on GMAC and outperforming the other solutions for $K_a \geq 225$.


\subsubsection{\textbf{\textcolor{black}{Random spreading-based schemes}}}
 In another design strategy for URA over GMAC, there is no slotting, and the active users utilize the entire transmission frame. In some works that follow this strategy, the idea is to spread the channel-coded bits throughout the transmission frame using random signatures. The random spreading idea was first introduced in \cite{pradhan2020polar}, where the message is divided into two parts: the first part determines the signature sequence, while the second part is encoded by a polar code. At the receiver, first, the selected signature sequences are detected by a correlation-based energy detector. Then, an iterative decoding algorithm consisting of MMSE filtering followed by single-user polar decoding is employed to recover the channel-coded bits. Moreover, SIC is applied at the end of each iteration. The general decoding process of random spreading-based solutions is illustrated in Fig. \ref{figdecspread}, \textcolor{black}{where $\mathbf{Y}$ is the received signal, $\mathbf{Y}_t$ is the residual of the received signal at the $t$-th iteration, and $D^t$, $\beta^t_i$, and $\hat{v}^t_i$ are the set of detected signature indices, LLR vector, and the message estimate of the $i$-th user at the $t$-th iteration, respectively.}

In \cite{pac}, the authors modify the scheme in \cite{pradhan2020polar} by replacing the polar codes with polarization-adjusted convolutional (PAC) codes, with a slight performance improvement and reduced implementation complexity. The performance of the scheme in \cite{pradhan2020polar} is improved in \cite{ahmadi2021random} by allowing users to randomly select a power level from a predefined set. This approach, called \textit{power diversity}, creates an artificial fading-like scenario, where the users with higher power levels are decoded first with a high probability, and those with lower power levels are decoded next after SIC. This results in a performance improvement over \cite{pradhan2020polar} for $K_a \geq 150$. In \cite{han2021sparse}, the authors propose to encode the data as the Kronecker product of two-component codewords, and employ an iterative decoder based on bilinear generalized AMP to decompose the Kronecker product and a soft-in-soft-out decoder for the individual components.  The authors in \cite{pradhan2021ldpc} replace the polar codes in \cite{pradhan2020polar} with LDPC codes to exploit the soft information produced by the belief propagation decoder to perform soft-input-soft-output (SISO) MMSE filtering. This approach enhances decoding performance by leveraging information from partially decoded codewords, resulting in the best performance among schemes based on random spreading. Moreover, enhanced spread spectrum ALOHA (E-SSA) \cite{herrero2008high}, an asynchronous RA protocol that combines ALOHA with direct sequence spread spectrum, is adapted for URA over GMAC in \cite{schiavone2024design}, where a wrap-around version of E-SSA is employed to adapt it for a synchronous URA setup. \textcolor{black}{In a recent work, random spreading with higher order pulse amplitude modulation (PAM) such as 8-PAM or 16-PAM is studied in \cite{Dang2025Higher}. The authors propose an AMP-based solution combined with expectation maximization and multiple measurement vectors, demonstrating that their proposed solution can support up to 1100 active users. This is the highest number of supported users reported in the literature for URA over GMAC for the practical PUPE value of $\epsilon=0.05$.}


\subsubsection{\textbf{\textcolor{black}{IDMA-based schemes}}}
In random spreading-based solutions, each active user occupies the entire frame by random spreading of channel-coded bits for data transmission. In some other works, active users distribute their codeword bits across the transmission frame in various ways. For instance, in the sparse interleave-division MA (IDMA) approach \cite{pradhan2022Sparse}, the message bits and the frame are divided into different parts. The users transmit a preamble sequence dictated by the first part of their message in the preamble part of the frame. The remaining bits are encoded using an LDPC code, zero-padded, and permuted by an interleaver determined by the preamble to be transmitted over the rest of the channel frame. This encoding process is illustrated in Fig. \ref{sparseIDMA}. At the receiver, after decoding the preamble part using the least absolute shrinkage and selection operator (LASSO), a single joint Tanner graph is employed for message recovery, leveraging the information on the location of transmitted symbols obtained from the identified preamble. \textcolor{black}{In a recent sparse IDMA-based scheme proposed in \cite{Yuan2025Unsourced}, the authors employ AMP for the recovery of the preamble part and convolutional coding in the data part, which improves the performance of \cite{pradhan2022Sparse} in high active user loads.}

A similar approach to the sparse IDMA in \cite{pradhan2022Sparse} is adopted in \cite{zheng2020polar}, employing polar codes; however, different code lengths are used among the users, and TIN is utilized as the decoding algorithm.  Moreover, in \cite{truhachev}, the message bits are first encoded by a high-rate error correction code. The encoded bits are divided into two parts, with the first part being mapped to a column of a common sensing matrix, which determines the interleaver and signature to be utilized in the second part after its repetition. The receiver utilizes AMP to recover the first part, followed by a multiuser decoding algorithm for the second part.

\textcolor{black}{While the works mentioned above consider the synchronous transmission, a fully asynchronous (i.e., with an unbounded time horizon) URA setup is studied in \cite{Karami2024}, where each user transmits a preamble (from a pool of available ones) for timing acquisition followed by its payload that is interleaved and multiplied by a signature similar to \cite{truhachev}. The receiver employs a sliding window decoding approach where correlation-based arrival time detection, multiuser decoding, and SIC operations are iteratively performed.}


\begin{table*}[t]
\footnotesize
\centering
\caption{A summary of URA solutions for GMAC}
\begin{tabular}{|m{.12\linewidth}|m{.12\linewidth}|m{.47\linewidth}| m{.08\linewidth}| m{.08\linewidth}|} 
 \hline
 \cellcolor[HTML]{BDBDBD} \centering Paper & \cellcolor[HTML]{BDBDBD} Approach & \cellcolor[HTML]{BDBDBD} \textcolor{black}{Description} & \cellcolor[HTML]{BDBDBD} \textcolor{black}{Channel Code} & \cellcolor[HTML]{BDBDBD} \textcolor{black}{Synchron. vs. Asynch.} \\ [0.5ex] 
 \hline
 \makecell{Polyanskiy, \\ 2017 \cite{polyanskiy2017perspective}}  & - & Information-theoretic formulation of URA, and Random coding achievability and converse bounds. & \textcolor{black}{-} & \textcolor{black}{-} \\ 
 \hline
\makecell{Ordentlich et al., \\ 2017 \cite{ordentlich}} & slotting & $T$-fold ALOHA-based concatenated coding scheme.  & \textcolor{black}{BCH} & \textcolor{black}{Synchron.}   \\
 \hline
 \makecell{Vem et al., \\ 2019 \cite{vem2019user}} &  slotting & $T$-fold IRSA-based scheme employing preamble transmission, channel coding, joint BP decoding, and SIC. &  \textcolor{black}{LDPC} & \textcolor{black}{Synchron.} \\
 \hline
 \makecell{Marshakov et al., \\ 2019 \cite{marshakov2019polar}} &  slotting & $T$-fold IRSA-based scheme employing preamble transmission, channel coding, joint successive cancellation decoding, and SIC. &  \textcolor{black}{Polar} & \textcolor{black}{Synchron.} \\
 \hline
  \makecell{Amalladinne et al., \\ 2020 \cite{Amalladinne2020Acoded}} &  CCS & Concatenated coding scheme employing inner CS and outer tree coding. & \textcolor{black}{-}  & \textcolor{black}{Synchron.}\\
   \hline 
  \makecell{Amalladinne et al., \\ 2022 \cite{comp4}} &  CCS & Concatenated coding scheme employing inner SPARC and outer tree codes. Interaction between inner AMP and outer BP decoding is proposed.  & \textcolor{black}{-} & \textcolor{black}{Synchron}. \\
 \hline
   \makecell{Nassaji et al., \\ 2024 \cite{nassaji2024dynamic}} &  CCS &  Concatenated coding scheme employing SPARCs and CRC encoding. Eliminating the outer tree code is proposed to improve the performance. & \textcolor{black}{-} & \textcolor{black}{Synchron.} \\
 \hline

    \makecell{Pradhan et al., \\ 2020 \cite{pradhan2020polar}} &  random spreading & Spreading-based scheme combining correlation-based signature detection, MMSE symbol estimation, single-user decoding, and SIC. & \textcolor{black}{Polar} & \textcolor{black}{Synchron.} \\
 \hline
     \makecell{Ahmadi et al., \\ 2020 \cite{ahmadi2021random}} &  random spreading & Spreading-based scheme employing similar operations with \cite{pradhan2020polar} and improve its performance with power diversity. & \textcolor{black}{Polar} & \textcolor{black}{Synchron.} \\
 \hline
    \makecell{Han et al., \\ 2021 \cite{han2021sparse}} &  random spreading & Spreading-based scheme combining Bayesian matrix factorization, BCJR decoding, and SIC at the receiver side. & \textcolor{black}{Convolutional} & \textcolor{black}{Synchron.} \\
 \hline
     \makecell{Pradhan et al., \\ 2021 \cite{pradhan2021ldpc}} &  random spreading & Spreading-based scheme combining energy detection, soft MMSE estimation, soft cancellation, single-user decoding, and SIC.  & \textcolor{black}{LDPC} & \textcolor{black}{Synchron.} \\
 \hline
    \makecell{Pradhan et al., \\ 2022 \cite{pradhan2022Sparse}} &  IDMA & Spreading-based scheme employing preamble transmission, joint Tanner graph decoding, and soft cancellation. & \textcolor{black}{LDPC} & \textcolor{black}{Synchron.} \\
 \hline
      \makecell{  \textcolor{black}{ Karami et al.,} \\ \textcolor{black}{ 2024 \cite{Karami2024} }  } & \textcolor{black}{IDMA} & \textcolor{black}{Spreading-based scheme employing correlation-based timing detection, multiuser decoding and SIC.} & - & \textcolor{black}{Asynch.} \\
 \hline
      \makecell{Yan et al., \\ 2023 \cite{odma}} &  ODMA & ODMA-based scheme employing blind pattern detection, joint graph decoding, and soft cancellation. & \textcolor{black}{RA} & \textcolor{black}{Synchron.} \\
 \hline
       \makecell{Ozates et al., \\ 2024 \cite{odma2}} &  ODMA & ODMA-based scheme employing energy test for pattern detection, single-user decoding, and SIC. & \textcolor{black}{Polar} & \textcolor{black}{Synchron.} \\
        \hline
        \makecell{Yan et al., \\ 2024 \cite{Yan2025Enhanced}} &  ODMA & ODMA-based scheme employing blind pattern detection, joint graph decoding, collision resolution, and soft cancellation. & \textcolor{black}{Non-binary LDPC} & \textcolor{black}{Synchron.} \\
        \hline   
       \makecell{ \textcolor{black}{ Ozates et al.,} \\ \textcolor{black}{ 2025  \cite{Ozates2025Fully}  } } & \textcolor{black}{ODMA} & \textcolor{black}{ODMA-based scheme employing nested sliding windows, timing detection by ODMA patterns, single-user decoding and SIC. } & \textcolor{black}{Polar}  & \textcolor{black}{Asynch.}  \\
        \hline        
\end{tabular}
\label{tablegmac}
\normalsize
\end{table*}


\subsubsection{\textbf{\textcolor{black}{ODMA-based schemes}}}
Another way to distribute the user codeword bits onto the transmission frame is proposed in \cite{odma}, where the authors employ on-off division MA (ODMA) \cite{odma1} in the context of URA. In this approach, the active users exploit a small disjoint portion of the transmission frame to transmit their data encoded by a repeat-accumulate code according to a randomly selected transmission pattern, with the remaining time instances being idle. At the receiver, the BS first recovers the transmission patterns and then employs multiuser detection and decoding over a sparse factor graph. The authors in \cite{odma2} investigate the ODMA technique in URA further, namely, they employ polar codes to improve the performance in the low active user load regimes combined with a simpler pattern detection algorithm. Another high-performing solution is proposed in \cite{Yan2025Enhanced}, which improves the scheme in \cite{odma} by employing a non-binary LDPC code. \textcolor{black}{Moreover, an analytical power division method that is formulated based on TIN-SIC is proposed in \cite{Zhang2025Sparse}, which achieves the same performance as that of \cite{Yan2025Enhanced} using 5G-NR LDPC codes instead of non-binary LDPC coding. Furthermore, the authors in \cite{Ozates2025Fully} examine the fully asynchronous setup introduced in \cite{Karami2024}, and propose a preamble-free ODMA-based solution. The receiver utilizes a sliding-window decoding approach with two nested windows. In the inner sliding window, an iterative decoder combining arrival time detection using the ODMA patterns, single-user polar decoding, and SIC is employed, while the outer window enhances the interference cancellation.} 


\begin{figure}[t]
    \centering
    \includegraphics[trim={.5cm 0cm 1cm 0.5cm},clip,width=1\linewidth]{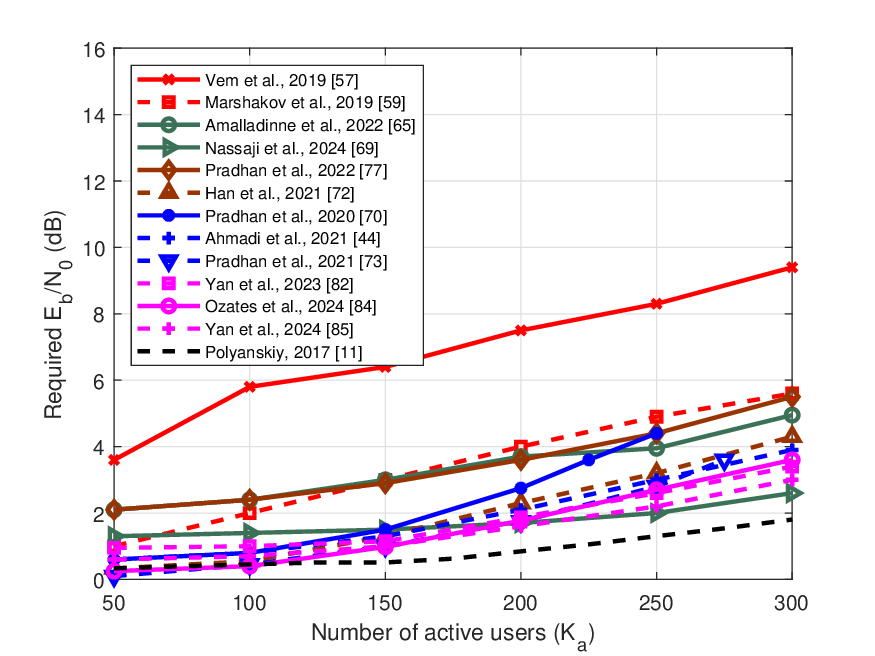}
    \caption{Performance comparison of the coding schemes in URA over GMAC for $n = 30000$, $B = 100$, and $\epsilon = 0.05$, which are the common set of parameters in the literature.}
    \label{figperfgmac}
\end{figure}

\subsection{\textcolor{black}{Summary and Comparison of GMAC Solutions }}
A summary of the URA solutions for GMAC is presented in Table \ref{tablegmac}, with comparisons among them in different aspects.

\subsubsection{\textbf{\textcolor{black}{Performance Comparison}}}
\textcolor{black}{The performance of the GMAC URA schemes is compared in terms of energy efficiency, the standard performance criterion in URA,} in Fig. \ref{figperfgmac} for $n = 30000$, $B = 100$, and $\epsilon = 0.05$, which are the common set of parameters in the literature. \textcolor{black}{Note that due to the vast number of papers, we have included only the state-of-the-art papers, at least in some regime, and those introducing a new approach when published.} The results in Fig. \ref{figperfgmac} demonstrate that random spreading-based schemes in \cite{pradhan2020polar,ahmadi2021random,pradhan2021ldpc} offer good performance for a low number of active users; however, their performance degrades with increasing the number of active users since each active user occupies the entire transmission frame. On the other hand, ODMA-based approaches in \cite{odma,odma2,Yan2025Enhanced} outperform the others, except for the solution proposed in \cite{nassaji2024dynamic}, in the high multiuser interference regime due to their sparsity. Meanwhile, they perform well in the low multiuser regime when combined with polar coding. Moreover, the CCS-based approaches can exhibit very good performance for a high number of active users when the rate loss due to the parity bits is alleviated, as shown in \cite{nassaji2024dynamic}. Compared to the achievability bound, the proposed schemes can perform similarly or slightly better for $K_a \leq 100$, and the gap is at most about 0.8 dB for $K_a \leq 300$.

\textcolor{black}{In summary, the performance trends over GMAC reveal a clear trade-off: While random spreading-based schemes are attractive at low user loads, their scalability is limited due to full-frame occupancy and high complexity. Conversely, ODMA-based approaches and recent CCS variants achieve robust performance in high-user regimes, benefiting from sparsity and efficient decoding. These results underscore that practical URA deployment must carefully balance energy efficiency, complexity, and user density, with CCS and ODMA schemes emerging as the most scalable options.}

\subsubsection{\textbf{Computational Complexity \textcolor{black}{Comparison}}}
\textcolor{black}{In terms of complexity, slotting is the most beneficial approach among the existing solutions for URA over GMAC, as the complexity of the operations, depending on the user count, increases super-linearly by the number of active users in the slot rather than $K_a$. In addition, slotting leads to smaller preamble/pilot lengths and a shorter channel code length $n_c$, thereby decreasing the complexity of the operations. Coded compressed sensing schemes also offer low complexity since slot-wise compressed sensing and low-complexity tree codes are employed in conjunction with each other. Specifically, its complexity order is linear or quadratic in $K_a$, depending on the regime \cite{Amalladinne2020Acoded}.  ODMA is another low-complexity approach, as only a small fraction of the frame is occupied by each user, which simplifies the iterative multiuser decoding process \cite{odma}. More specifically, its complexity order depends linearly on $n_c$ and $K_a$. On the other hand, random spreading-based schemes where each user occupies the entire frame have the highest complexity, as the complexity order is cubic in $K_a$ due to the MMSE filtering \cite{pradhan2020polar}. 
}

\section{URA over Fading Channels}

\textcolor{black}{GMAC is an ideal channel model that does not capture many practical effects of the communication channel.
With this motivation, more realistic fading MAC models, where the transmitted signals of the users are attenuated with random fading coefficients, are considered in several works in the URA literature. }

\textcolor{black}{There are some challenges in extending and applying the URA over GMAC solutions to fading scenarios. The primary challenge in fading MAC is estimating the channel coefficients at the receiver side and designing the decoder to account for fading. In the literature, two approaches are proposed to address this problem; the first is to embed the channel estimation into the decoding algorithm without requiring a separate channel estimation step (e.g., \cite{kowshik2020energy, andreev2020polar, decurninge2020tensor}). The second approach is to exploit a part of the frame for pilot transmission, perform explicit channel estimation using the received signal in the pilot part, and utilize the estimated coefficients for detection in the data part (e.g., \cite{fading4,fengler2022pilot,Gkagkos2023FASURA}). While most schemes developed for single-antenna fading channels, and the CCS-based and tensor-based schemes in URA over MIMO fading channels opt for the former, there is also a vast literature on pilot-based schemes. The main algorithms applied for channel estimation in pilot-based schemes are MMSE \cite{odma2, fengler2022pilot, Gkagkos2023FASURA, Ozates2023Aslotted,nassaji,Ozates2024ODMA} and AMP \cite{fading4, Su2022Massive,Su2023Index}. Note that in most URA works over fading channels, quasi-static fading is considered.} 



In this section, we first review the bounds on the performance of URA solutions taking into account channel fading that have been provided in the literature. We then cover the URA solutions that present low-complexity coding schemes to achieve those bounds. To this end, in order to explain the URA approaches that take into account channel fading in a simpler setup for better understanding, we first cover the single-antenna fading solutions and then turn our attention to the MIMO fading ones.

\subsection{\color{black}Performance Bounds}
In \cite{andreev2020polar,kowshik2019quasi,kowshik2021fundamental,kowshik2020energy}, the authors derive some performance bounds for the URA framework over a Rayleigh fading channel. Specifically, the authors in \cite{kowshik2020energy} present a converse bound based on the results in \cite{simo} and the meta-converse bound in \cite{polyanskiythesis}. These results numerically demonstrate that the $E_b/N_0$ requirements for a fading channel are necessarily higher than those for the AWGN channel. They also argue that, for typical system parameters in this setup, a Fano-type converse bound is worse than the proposed one. 




In \cite{kowshik2019quasi} and \cite{kowshik2021fundamental}, achievability and converse bounds are proposed for the cases of known and unknown CSI in an asymptotic setting, where the number of users grows linearly with the blocklength. The authors in \cite{kowshik2020energy} provide both asymptotic and non-asymptotic achievability and converse bounds for the case of unknown CSI. In \cite{andreev2020polar}, authors derive an achievability bound for the URA scheme over the Rayleigh fading channel by considering a slotted structure and perfect SIC in their analysis, which is tighter than the other bounds.
However, since the channel coefficients used in an SIC process are estimated, hence imperfect, the assumption of a perfect SIC is not realistic, which makes the result in \cite{andreev2020polar} an approximation rather than an achievability one.

Some achievability and converse bounds for URA over MIMO Rayleigh fading channels are obtained in \cite{Gao2023Energy,Gao2024}. Among the developed results, the ML-based achievability bound is obtained by employing a projection decoder developed in \cite{Andreev2024Unsourced}. \textcolor{black}{Moreover, the results in \cite{Gao2023Energy} are extended to the case of random and unknown number of active users in \cite{Gao2025Unsourced}}.


\subsection{Single-Antenna Fading Channels}
We first examine the URA solutions over fading channels that assume a single-antenna receiver. The works in \cite{pradhan2022Sparse,fading4,Yun2024Erasure,andreev2020polar,kowshik2020energy,Andreev2022cod,andreev2019low, Ozates2023Unsourced,Amalladinne2019asy, Kowshik2019sho} introduce practical coding schemes tailored for this setup. Among these works, some assume that the received signals from all users are synchronized \cite{pradhan2022Sparse,andreev2020polar,kowshik2020energy,   Andreev2022cod, fading4, Yun2024Erasure}, while others take into account asynchronous transmission \cite{Ozates2023Unsourced,andreev2019low, Amalladinne2019asy, Kowshik2019sho}. To estimate the delay of asynchronous signals, \cite{Ozates2023Unsourced,andreev2019low, Kowshik2019sho} use OFDM-based transmission to convert the time delay to a phase shift in the frequency domain, while the authors in \cite{Amalladinne2019asy} consider time-domain transmission where the delay of a signal is detected via a correlation-based approach.

Focusing on Rayleigh fading channels, we present existing solutions for URA over fading channels, categorized based on their general approach.

\subsubsection{\textbf{CCS-based schemes}}
As explained in the previous section, CCS was initially proposed for URA over GMAC. In subsequent literature, it has also been applied to the fading scenario, see \cite{Andreev2022cod,Amalladinne2019asy,Yun2024Erasure}. 
In \cite{Andreev2022cod}, the authors \textcolor{black}{replace the tree code} of the GMAC CCS scheme in \cite{Amalladinne2020Acoded} by \textcolor{black}{list recoverable codes that are} capable of correcting $t$ errors. \textcolor{black}{They propose two outer code constructions}: The first is a modified tree code with the ability to recover $t$ errors (called the $t$-tree), and the second is a Reed-Solomon (RS)-based code. In \cite{Yun2024Erasure}, the outer encoded message bits are partitioned into sub-blocks, and each sub-block is transmitted in a different slot after CS-type encoding, with AMP at the receiver. Moreover, in \cite{Amalladinne2019asy}, the authors utilize CCS for the asynchronous URA model. More specifically, during the encoding stage, they append $T_z$ additional zeros to the end of each CS sequence, where $T_z$ represents the maximum possible delay of the system. At the receiving end, the decoder takes into account all possible delays, ranging from one symbol to a maximum of $T_z$-symbol delays during the decoding process.  


\subsubsection{\textbf{Joint decoding-based schemes}}
\begin{figure}
	\centering
\includegraphics[width=.9\linewidth]{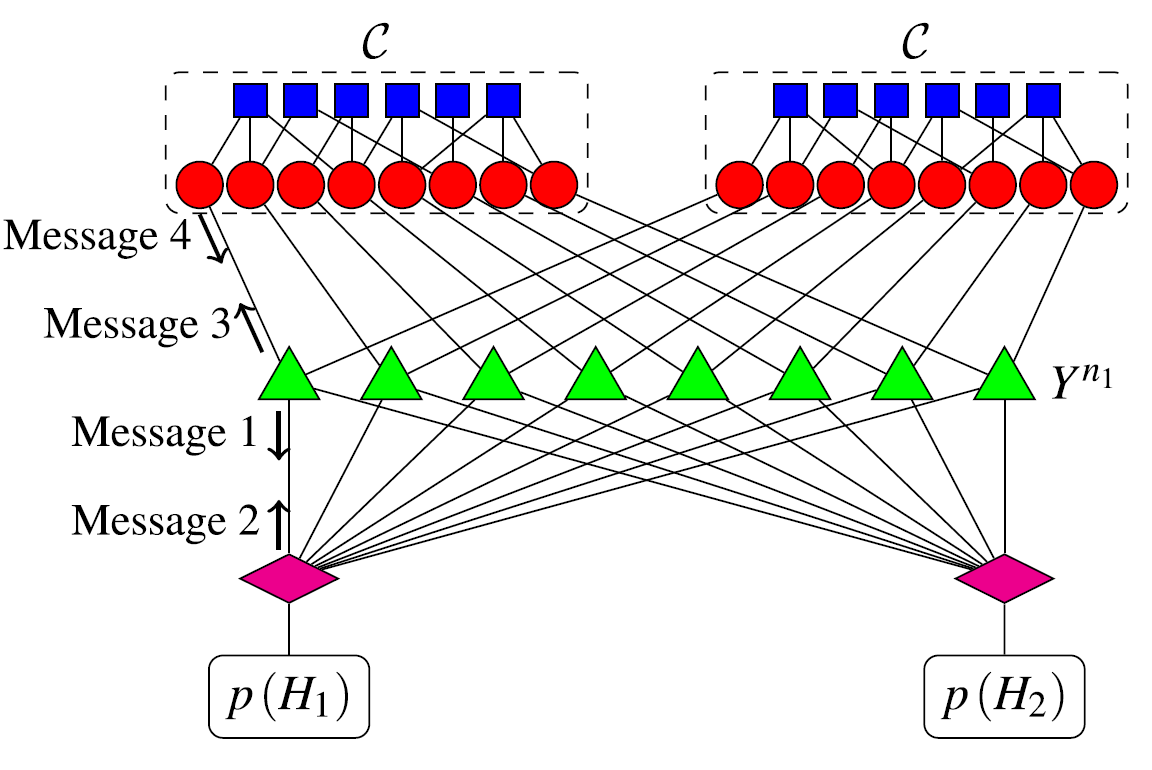}
		\caption{ Iterative BP decoder for a two-user LDPC-based scheme with joint decoding \cite{kowshik2020energy}.}	\label{fig_BeliefProp}
	\end{figure}

In \cite{kowshik2020energy}, a joint decoding approach is employed. For encoding, this scheme partitions the frame into distinct slots. Each user transmits its LDPC codeword through a randomly chosen slot. As previously mentioned, slot allocation offers the significant benefit of mitigating the per-slot interference by distributing users across slots.
This scheme utilizes an iterative belief propagation (BP) decoder, as illustrated in Fig. \ref{fig_BeliefProp}. The graph incorporates four distinct node types: variable nodes (in red); check nodes (in blue), which constitute the traditional LDPC decoder; \textit{functional} nodes (in green) corresponding to the symbols of the received signal; and a fourth type (in magenta) corresponding to the fading coefficients. This graph allows them to update the estimated values of both the channel coefficients and the symbols simultaneously for each user.

\subsubsection{\textbf{TIN-SIC-based schemes}}
A lower-complexity alternative to joint decoding is the TIN-based decoding. Similar to GMAC, the TIN solution for fading channels operates under the assumption that a single user is present and treats the signals of other interfering users as noise. Consequently, this approach results in an exceptionally low average SNR because it encounters interference-plus-noise as the system's overall noise level, leading to inferior performance. However, when combined with an appropriate channel code and SIC, the performance can be improved.

\begin{table*}[t]
\footnotesize
\centering
\caption{A summary of URA solutions for fading channels with a single-antenna receiver.}
\begin{tabular}{|m{.13\linewidth}|m{.06\linewidth}|m{.45\linewidth}| m{.07\linewidth}|m{.06\linewidth}|m{.05\linewidth}|}
\hline
\cellcolor[HTML]{BDBDBD} \centering Paper &\cellcolor[HTML]{BDBDBD}Approach & \cellcolor[HTML]{BDBDBD}Description& \cellcolor[HTML]{BDBDBD}\makecell{Synchron.\\vs. Asynch.} & \cellcolor[HTML]{BDBDBD} \makecell{\textcolor{black}{Channel}\\\textcolor{black}{Estimation}} & \cellcolor[HTML]{BDBDBD} \textcolor{black}{Channel Code}  \\
\hline
\makecell{Kowshik et al.,\\ 2019\&2021 \cite{kowshik2019quasi,kowshik2021fundamental} } & -& Achievable and converse bounds for known and unknown CSI in an asymptotic setting + comparing with CDMA, TDMA, and FDMA.& Synchron. & \textcolor{black}{-} & \textcolor{black}{-} \\ \hline
\makecell{Kowshik et al.,\\ 2019 \cite{Kowshik2019sho} } & TIN-SIC &  $T$-fold ALOHA-based scheme employing OFDM, channel coding, and SIC.  & Asynch. & \textcolor{black}{Embedded} & \textcolor{black}{LDPC}\\\hline
\makecell{Andreev et al.,\\ 2020 \cite{andreev2020polar} } &TIN-SIC& $T$-fold ALOHA-based scheme employing channel coding, single-user decoding, and SIC.& Synchron. & \textcolor{black}{OMP} & \textcolor{black}{Polar}  \\\hline
\makecell{Kowshik et al.,\\ 2020 \cite{kowshik2020energy} } &\makecell{joint \\decoding}&  Achievability and converse bounds for known and unknown CSI + a scheme employing slotting, channel coding, and joint BP decoding.  & Synchron. & \textcolor{black}{Embedded} & \textcolor{black}{LDPC}\\\hline
\makecell{Amalladinne et al.,\\ 2019 \cite{Amalladinne2019asy} } & CCS &  Extension of CCS in \cite{Amalladinne2020Acoded} to accommodate asynchronous scenarios. & Asynch. & \textcolor{black}{Embedded} & \textcolor{black}{-}\\\hline
\makecell{Andreev et al.,\\ 2022 \cite{Andreev2022cod} } & CCS &  Improving CCS in \cite{Amalladinne2020Acoded} by replacing the outer tree code with 1) a code capable of correcting $t$ errors, and 2) a Reed–Solomon code.  & Synchron. & \textcolor{black}{Embedded} & \textcolor{black}{RS}\\\hline
\makecell{Yun et al.,\\ 2024 \cite{Yun2024Erasure} } & CCS &  A slotted scheme employing CCS and AMP in a block fading model.  & Synchron. & \textcolor{black}{Embedded} & \textcolor{black}{-} \ \\\hline
\makecell{Nassaji et al.,\\ 2022 \cite{fading4} } & Spreading &  A non-slotted scheme employing preamble transmission, channel coding, repetition, permutation, scrambling, and SIC.  & Synchron.  & \textcolor{black}{AMP} & \textcolor{black}{Polar}\\\hline
\makecell{\textcolor{black}{Ozates et al.,} \\ \textcolor{black}{2024 \cite{odma2}}} &  \textcolor{black}{ODMA} & \textcolor{black}{ODMA-based scheme employing gOMP for pilot and pattern detection, single-user decoding and SIC.}  & \textcolor{black}{Synchron.}  & \textcolor{black}{LMMSE} & \textcolor{black}{Polar} \\
    \hline
\makecell{Ozates et al.,\\ 2023 \cite{Ozates2023Unsourced} } & OFDM+ SIC &   A slotted scheme in the frequency-selective channel employing OFDM, pilot transmission, channel coding, and SIC.& Asynch. & \textcolor{black}{OMP} & \textcolor{black}{Polar}\\\hline
\end{tabular}
\label{Table_SingleAntenna}
\normalsize
\end{table*}

\textbf{TIN-SIC with LDPC coding:} In \cite{andreev2019low}, building upon the synchronous schemes presented in \cite{kowshik2020energy}, the authors present a low-complexity coding scheme which incorporates OFDM techniques to mitigate the effects of the asynchronous transmission by converting the time delays into phase shifts, and employing SIC to decode the user packets. Further numerical results for this scheme are presented in \cite{Kowshik2019sho}.

\begin{figure}
    \centering
        \includegraphics[width=1\linewidth]{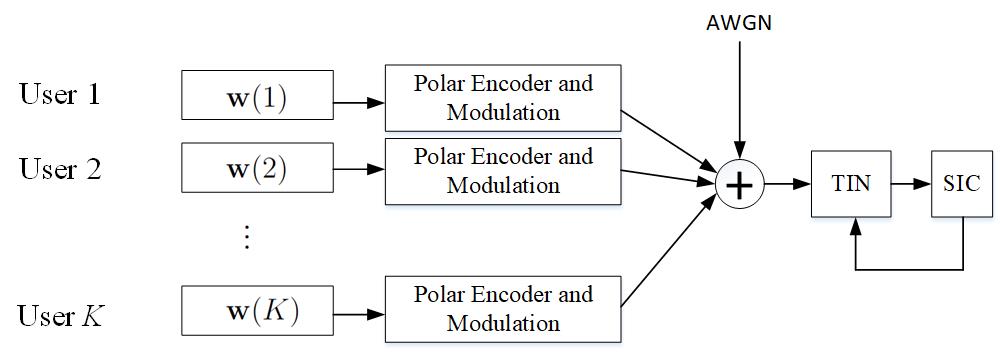}
	\caption{ Encoding and decoding processes of TIN-SIC scheme with polar codes in \cite{andreev2020polar} in a slot with $K$ users.}	\label{Fig_tinSIC}
\end{figure}

\textbf{TIN-SIC with Polar Coding:} In \cite{andreev2020polar}, a TIN-SIC solution in conjunction with polar coding is presented, where the frame in this approach is divided into $S$ slots. Each user randomly selects a slot to transmit its polar-coded and modulated signal. For decoding, the scheme considers a TIN strategy followed by SIC. Since different users in a slot experience different fading coefficients, some of them (those with larger fading coefficient amplitudes) are more likely to be decoded by TIN. After decoding the strongest user's signal while treating the others as noise, the channel coefficient is estimated using orthogonal matching pursuit (OMP). Given the estimated channel coefficient and the decoded signal of a user, its contribution can be removed from the received signal using SIC. This strategy continues until no signal is successfully detected during an iteration. It is important to emphasize that the determination of whether a decoded message is successful relies on the use of a cyclic redundancy check (CRC) message sequence. The encoding and decoding process of the scheme with polar coding with SIC in a slot with $K$ users is illustrated in Fig. \ref{Fig_tinSIC}.

\subsubsection{\textbf{Alternative solutions}}
Some other solutions also focus on URA over fading channels with a single-antenna receiver. For instance, in \cite{fading4}, the active users utilize the entire transmission frame to transmit their data. Specifically, the frame is divided into pilot and data parts. The active users select a pilot sequence from a common codebook for transmission in the pilot part, and they repeat their codeword bits for the remainder of the transmission time frame, with permutation and scrambling applied in the data part. At the receiver, AMP is employed for joint pilot detection and channel estimation, and the data part is recovered by iterative data estimation and interference cancellation using the repeated codeword bits.

\textcolor{black}{The ODMA approach is also applied to the single-antenna fading scenario in \cite{odma2} by allocating a portion of the frame for pilot transmission while a polar codeword is transmitted in the rest of the frame by ODMA. At the receiver side, the selected pilots and patterns are detected by a generalized OMP (gOMP) algorithm, followed by linear MMSE (LMMSE) channel estimation, single-user decoding, and SIC to recover the transmitted messages.}
Furthermore, the authors in \cite{Ozates2023Unsourced} study URA over frequency-selective channels and propose OFDM to mitigate adverse channel effects combined with CS-based activity detection and TIN-SIC for message recovery at the receiver side.

\begin{figure}[t]
	\centering
\includegraphics[trim={.5cm 0cm 1cm 0.5cm},clip,width=1\linewidth]{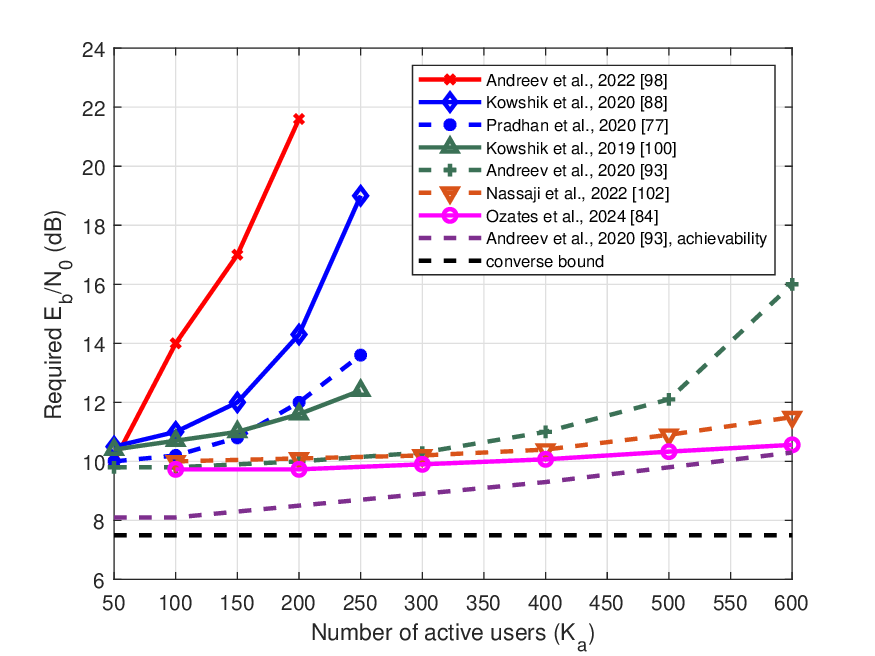}
		\caption{ Performance comparison of the coding schemes for URA over fading channel with a single-antenna receiver, with a target PUPE of $0.1$, $n=30000$, and $B=100$ (the common set of parameters in the literature).}	
  \label{Fig_Preliminary_Fading}
\end{figure}

\subsection{\textbf{\textcolor{black}{Summary and Comparison of Single-Antenna Fading Solutions}}} 

A summary of URA studies over fading channels involving a single-antenna receiver is given in Table \ref{Table_SingleAntenna}, \textcolor{black}{where \textit{Embedded} under channel estimation column refers to the case that estimation of the fading coefficient is included as a part of the decoding algorithm.}

\subsubsection{\textbf{\textcolor{black}{Performance Comparison:}}}
The performance of different coding schemes developed for URA over single-antenna fading channels is compared in Fig. \ref{Fig_Preliminary_Fading} assuming $n = 30000$, $B = 100$, and $\epsilon = 0.1$. 
{\color{black}The reason that some diagrams seem not to be fully plotted is that those solutions do not work after a certain active user load, i.e., they cannot achieve the target PUPE, no matter how much the $E_b/N_0$ is increased.}
As can be observed from Fig. \ref{Fig_Preliminary_Fading}, the CCS-based scheme in \cite{Andreev2022cod} exhibits lower energy efficiency compared to the other alternatives, as it employs a suboptimal coding scheme, whereas the others use strong channel codes, such as LDPC or polar codes. The performance of the LDPC-based scheme with joint decoding in \cite{kowshik2020energy} is inferior to that of the ones in \cite{pradhan2022Sparse}, \cite{Kowshik2019sho}, and \cite{andreev2020polar}, which can be attributed to the absence of an efficient SIC block in this LDPC-based approach. On the other hand, the TIN-SIC approach in \cite{Kowshik2019sho} and \cite{andreev2020polar} offers a better performance as it allows the receiver to recover the messages of users with a stronger fading coefficient, subtract their effects, and decode the remaining users in subsequent iterations, especially when combined with polar coding as in \cite{andreev2020polar}. We also note that the TIN-SIC scheme in \cite{andreev2020polar} is outperformed by the spreading-based solution in \cite{fading4}, and the ODMA-based scheme employing polar codes in \cite{odma2}, which offers the best performance among the proposed schemes, \textcolor{black}{due to the sparsity of the transmissions}.

\textcolor{black}{Overall, the results in fading channels show that CCS-based solutions lack a strong channel code to provide a high energy efficiency. Simple TIN-SIC solutions offer a good performance in low user loads, however, they demonstrate the limitations of weak interference management as device density grows. The comparative analysis suggests that practical single-antenna fading URA systems will favor ODMA and spreading-based approaches, which can be flexibly adapted to varying user loads without incurring excessive complexity.}

\subsubsection{\textbf{Computational Complexity \textcolor{black}{Comparison}}}
{\color{black} We now compare the computational complexity of the URA solutions over single-antenna fading channels. The tree-code-based CCS approach has a low complexity, which is linear in the size of the list of messages at each decoding step and exponential in the number of message bits in each slot \cite{Andreev2022cod}. In addition, the decoding operations in \cite{Andreev2022cod} can be carried out by additions, which are easier to implement on hardware compared to multiplications. The complexity of the $T$-fold ALOHA-based schemes employing LDPC with joint decoding \cite{kowshik2020energy} or polar coding with TIN-SIC \cite{andreev2020polar} is mainly determined by the parameter $T$, that is, the number of users that the decoder tries to recover in each slot. Note that TIN-SIC is more preferable than joint decoding since it utilizes single-user decoding, which makes it generally less complex and more parallelizable. Spreading-based and ODMA-based schemes, proposed in \cite{fading4} and \cite{odma2}, respectively, have higher complexities compared to their counterparts in exchange for their superior performance. For instance, the complexity of MMSE filtering for fading coefficient estimation in \cite{odma2} is cubic in $K_a$.}

\subsection{MIMO Fading Channels}
 We now turn our focus to the URA solutions developed for MIMO fading channels, which can be categorized into five groups: CCS-based, tensor-based, pilot-based, ODMA-based, and Bayesian approaches, with inevitable overlaps among them. 
\begin{figure*}[t]
\centering
\includegraphics[width=.7\linewidth]{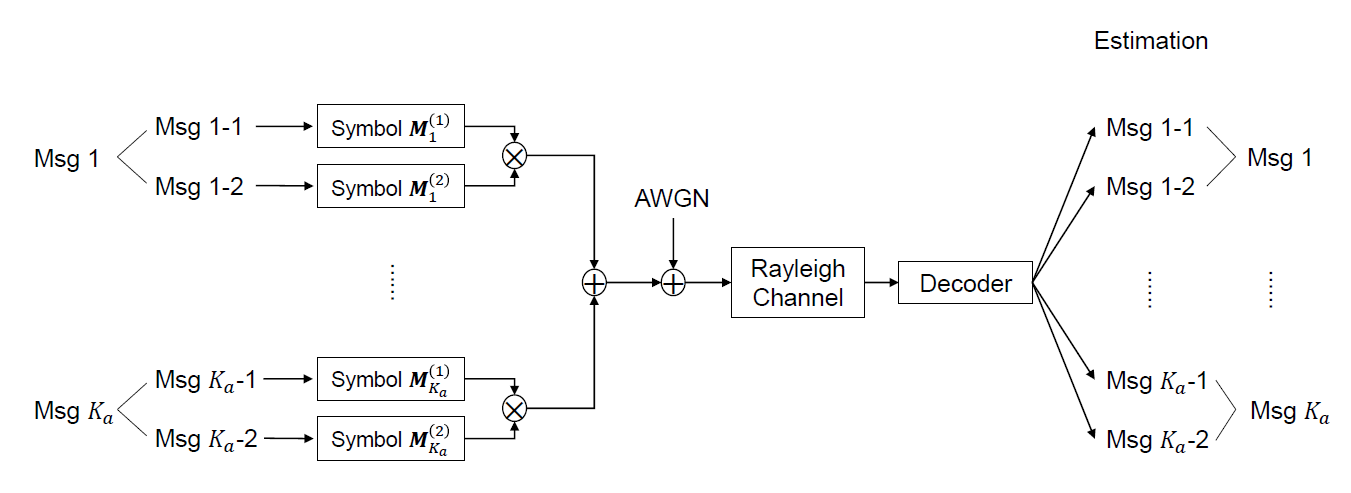}
\caption{ Encoding and decoding structures of tensor-based schemes for $d=2$
\cite{Luan2022Modulation}. }	
\label{Tensor_fig}
\end{figure*}

\subsubsection{\textbf{\textcolor{black}{CCS-based schemes}}}

In CCS-based algorithms for URA over MIMO fading channels \cite{fengler2021non,Shyianov2021Massive}, the bit sequence of each user is divided into $S$ sub-messages, which are then mapped to codewords in pilot codebooks with lower dimensions to obtain $S$ different pilots, transmitted through $S$ different slots. At the receiving end, $K_a$ sub-messages are detected within each slot using the ML \cite{fengler2021non} or hybrid generalized approximate message passing (HyGAMP) \cite{Shyianov2021Massive} technique, and they are then combined to form the complete length-$B$ bit sequence for each user. To establish connections among these sub-messages, the authors in \cite{fengler2021non} incorporate parity-check bits into each sub-message, while \cite{Shyianov2021Massive} relies on the correlation between channel coefficients of each user in different slots (assuming a quasi-fading channel model). Eliminating the need for additional parity bits enhances the performance of \cite{Shyianov2021Massive} compared to \cite{fengler2021non}. 

Eliminating the parity bits in \cite{fengler2021non} turns the problem into an uncoupled compressed sensing one, which is also considered in some recent works \cite{Tian2024Design,Agostini2024BiSPARC,Tian2024Exploiting,Zhang2024Blind}. To stitch the decoded sub-messages, the authors in \cite{Tian2024Design} employ a joint Bayesian decoding approach that exploits the channel characteristics, while geometric similarities between the codeword covariance matrices are utilized in the decoding algorithm in \cite{Tian2024Exploiting}, and the similarity of the estimated channel vectors is used in \cite{Zhang2024Blind}. The structure of the inner SPARC code allows parallel single-user outer decoding of user messages in \cite{Agostini2024BiSPARC}, which mitigates the need for the parity check bits. An uncoupled structure is employed in \cite{Xie2024Massive} to design a CCS-based URA scheme for near-field communications. Furthermore, a two-phase CS coding scheme is proposed in \cite{Jiang2024Two}, where a long sub-block recovered in the first phase is used for channel estimation, and very short-length sub-blocks are recovered in the second phase by employing a mutually orthogonal codebook and a hybrid AMP in an uncoupled structure. Moreover, the heterogeneous mobility of the massive number of users is considered in \cite{Ruan2023Unsourced } via a two-codebook aided URA scheme with covariance-based maximum likelihood detection (CB-MLD) at the receiver, where the static users employ a larger codebook while the mobile users utilize a smaller one in both the codeword length and the number of codewords. \textcolor{black}{Moreover, an uncoupled CS scheme employing polar coding and power diversity in combination with AMP-based decoding and SIC at the receiver side is proposed in \cite{Zhang2024Unsourced}.}

\subsubsection{\textbf{\textcolor{black}{Tensor-based schemes}}}
Another trend in MIMO URA is the tensor-based schemes \cite{decurninge2020tensor,mimo4,Luan2022Modulation}, where each user sends a rank-1 tensor of order $d$. Therefore, the received signal is a rank-$K_a$ tensor of order $d+1$ (considering the channel coefficient vector as an extra dimension of the tensor). For decoding, they conduct tensor decomposition on the received signal, followed by de-quantization of the tensors. In particular, in \cite{decurninge2020tensor}, by tensor decomposition, the rank-$K_a$ tensor of order $d+1$ is decomposed into $K_a$ rank-1 tensors of order $d+1$, which is equivalent to separating the signals of $K_a$ users, where the transmitted tensors can be identified by the first $d$ dimensions, and the $(d+1)$th dimension gives the channel vector of the user. Then, by de-quantizing the tensor of each user, its transmitted bit sequence is detected. By employing OFDM transmission, this scheme is extended to the case with timing offsets in \cite{mimo4}. On the other hand, the authors in \cite{Luan2022Modulation} propose a block-term decomposition for the decomposition of 3-dimensional tensors, where two low-rank factor matrices and a factor vector are utilized in each block term. \textcolor{black}{Moreover, the authors in \cite{Fang2025Polar} study tensor-based modulation with soft decoding, and propose a polar-coded tensor-based scheme with iterative Bayesian decoding with feedback at the receiver side.} The overall coding structure of tensor-based methods is depicted in Fig. \ref{Tensor_fig}.

\begin{figure}[t]
\centering
\includegraphics[width=1\linewidth]{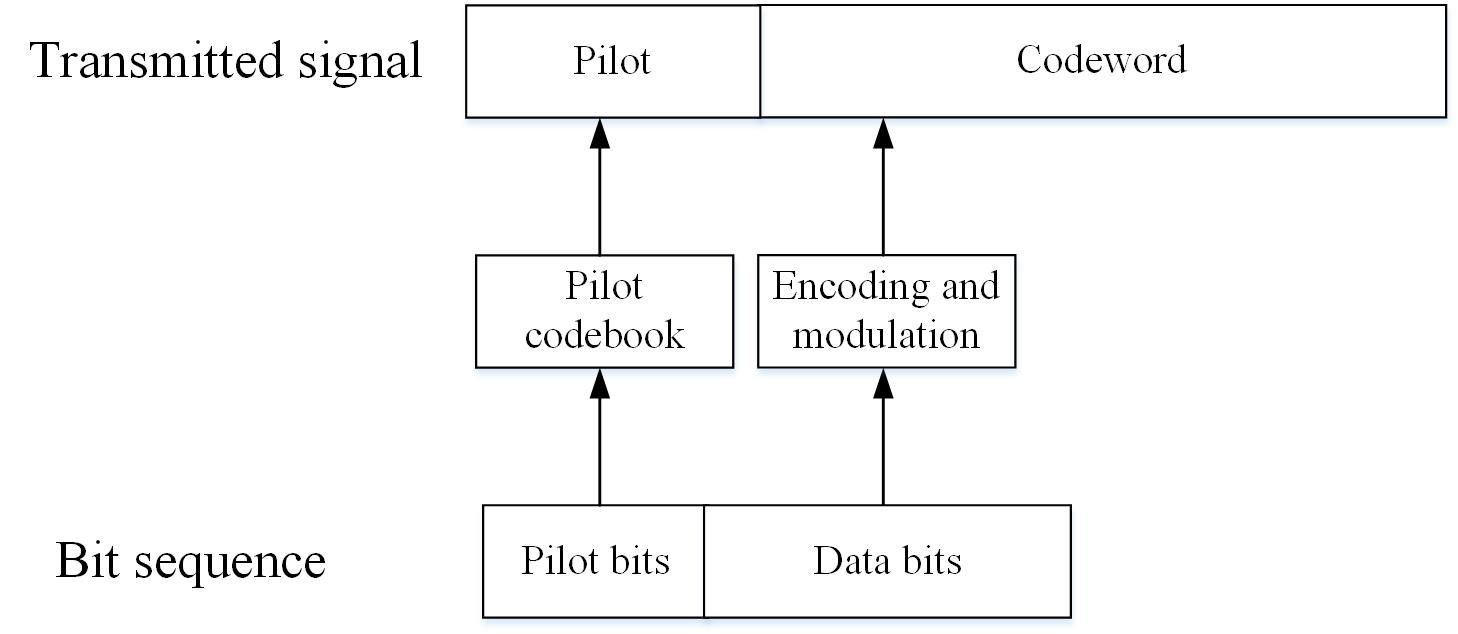}
\caption{ Encoding structure of pilot-based schemes. }
\label{PilotBased_encoder}
\end{figure}

\begin{figure}[t]
\centering
   \includegraphics[trim={0.5cm 10cm 10cm 0.5cm},clip,width=1\linewidth]{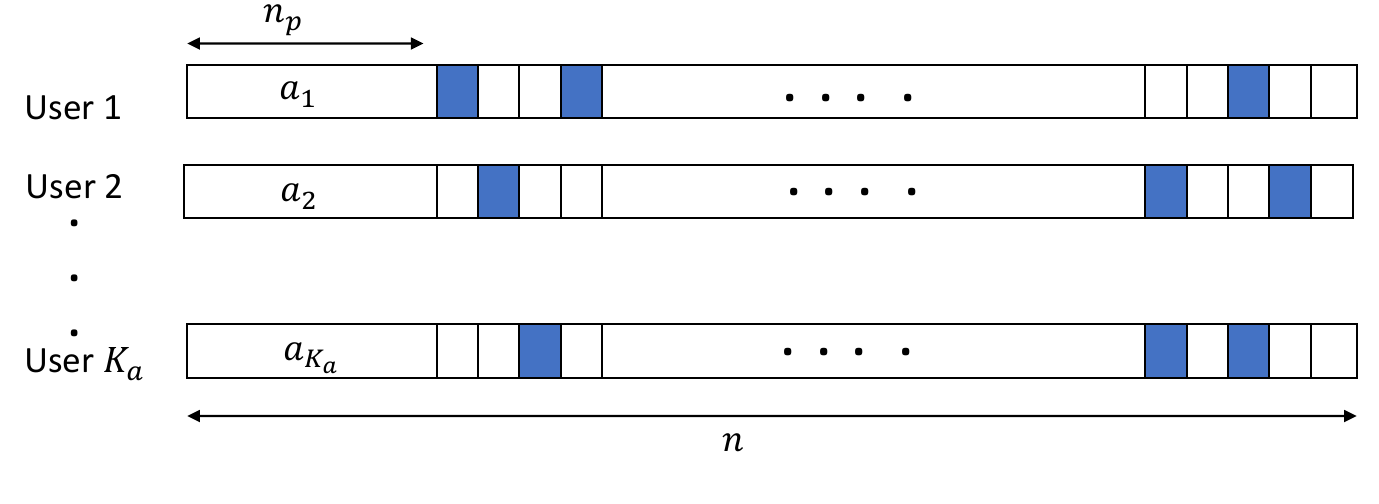}
\caption{ An illustration of the ODMA idea for the fading scenario with the pilot and data parts. Colored boxes show the utilized time instances in the data part.}
\label{odmafading}
\end{figure}

\subsubsection{\textbf{\textcolor{black}{Pilot-based schemes}}}
\begin{figure*}[h]
\centering
\includegraphics[width=.98\linewidth]{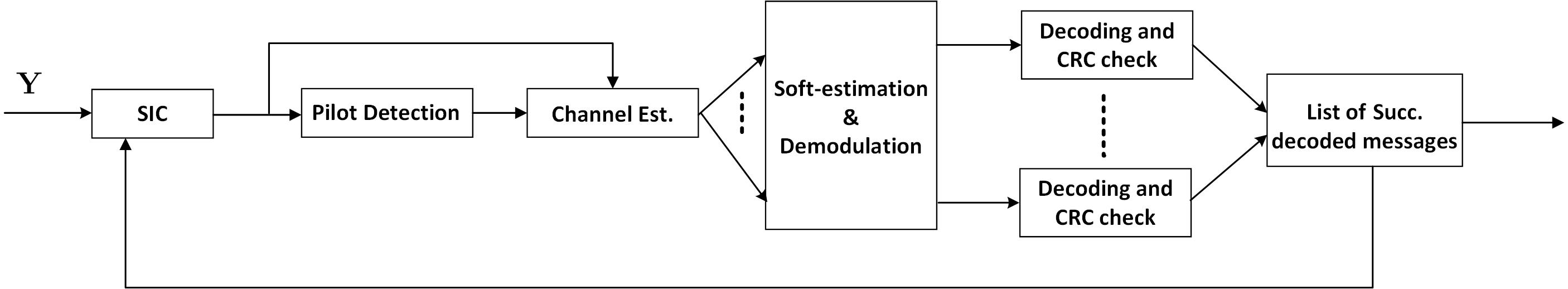}
\caption{ Decoding structure of pilot-based schemes. }	
\label{PilotBased_decoder}
\end{figure*}

Pilot-based techniques constitute another category of algorithms in the context of MIMO URA systems, where the idea is to divide the user bit sequences into pilot and data parts. In the existing solutions, the transmitted signal of each user also consists of two parts, as depicted in Fig. \ref{PilotBased_encoder}: the pilot part which is obtained by mapping the pilot bits to a codeword in a pilot codebook (a non-orthogonal codebook \cite{fengler2022pilot,Gkagkos2023FASURA,Su2022Massive,Su2023Index,Ozates2023Aslotted,Ozates2024ODMA}, or a codebook consisting of multiple stages of orthogonal pilots \cite{Ahmadi2023Unsourced}), and the data part which is encoded by a channel code (a polar \cite{Ahmadi2023Unsourced,fengler2022pilot,Gkagkos2023FASURA,Ozates2023Aslotted} or an LDPC code \cite{Su2022Massive,Su2023Index}). For decoding, as shown in Fig. \ref{PilotBased_decoder}, these schemes iteratively employ the following steps: 

\begin{itemize}
    \item At the initial stage, the receiver detects the active pilots using a user activity detection technique, such as gOMP \cite{Ozates2023Aslotted,Ozates2024ODMA}, AMP \cite{Su2023Index,nassaji}, multiple-measurement vector AMP (MMV-AMP) \cite{fengler2022pilot}, energy detector \cite{Gkagkos2023FASURA}, NP hypothesis testing \cite{Ahmadi2023Unsourced}, or CB-MLD \cite{Su2022Massive}. \textcolor{black}{Moreover, an improved version of the NNLS algorithm exploiting multiple measurements for activity detection is proposed in \cite{Sun2024Non}}.
    \item The channel vectors corresponding to each detected pilot are estimated by techniques such as MMSE \cite{fengler2022pilot, Gkagkos2023FASURA, Ozates2023Aslotted,nassaji,Ozates2024ODMA}, MRC \cite{Ahmadi2023Unsourced} or AMP \cite{Su2022Massive,Su2023Index}.
    \item Using the estimated channel coefficients, soft estimates of the transmitted codewords are obtained via decoupling techniques, such as MRC \cite{fengler2022pilot, Ahmadi2023Unsourced, Ozates2023Aslotted,Ozates2024ODMA}, MMSE \cite{Gkagkos2023FASURA} or AMP \cite{Su2022Massive}, which are then used to generate log-likelihood ratio (LLR) values to be fed to single-user decoders \cite{Ahmadi2023Unsourced,fengler2022pilot, Gkagkos2023FASURA, Su2022Massive, Ozates2023Aslotted,nassaji} or a joint factor graph \cite{Su2023Index}.
    \item In most pilot-based schemes, following the SIC approach, the contribution of the successfully decoded codewords is removed from the received signal to decrease the interference level at the subsequent iterations. Note that in \cite{Ahmadi2023Unsourced,Gkagkos2023FASURA,Ozates2023Aslotted,Ozates2024ODMA}, the channel vectors are re-estimated using the successfully decoded codewords for improved interference cancellation at the SIC stage. 
\end{itemize}

\subsubsection{\textbf{\color{black}ODMA-based schemes}}
ODMA is a powerful technique that was originally proposed for URA over GMAC. However, it is also extended to the fading scenario for the case of a single-antenna receiver in \cite{odma2} and a massive MIMO receiver in \cite{Ozates2024ODMA, Hao2024Superimposed, Zhang2024Turbo, Zhang2025Probabilistic, Zhang2024Efficient, Zhang2024Dictionary, Zhang2025LIM}. The main idea \textcolor{black}{in \cite{Ozates2024ODMA}}  is to allocate a part of the frame for pilot transmission, which facilitates the channel estimation at the receiver side, while the rest of the frame is utilized for data transmission in an ODMA manner as illustrated in Fig. \ref{odmafading}. \textcolor{black}{At the receiver side, the gOMP algorithm is employed to detect the active pilots, followed by LMMSE channel estimation, MRC symbol estimation, single-user polar decoding, and SIC to recover the data part.} Note that if the channel is a GMAC, the entire frame is used for data transmission. On the other hand, the authors in \cite{Hao2024Superimposed} consider superimposed pilot and data transmission, where the pilot sequence occupies the entire frame and the data signal is transmitted using ODMA. The receiver operation begins with energy detection-aided non-Bayesian pilot detection, and the same steps as in \cite{Ozates2024ODMA} are used for decoding the data part. A slotted ODMA-based scheme combining preamble transmission and convolutional LDPC coding is proposed in \cite{Zhang2024Turbo}, with AMP decoding for the preamble and turbo probabilistic data association for multiuser detection of the data part at the receiver side. \textcolor{black}{In another slotted ODMA-based scheme \cite{Zhang2025Probabilistic}, the authors consider uncoupled ODMA where the pilot sequences and transmission patterns are selected by different bits of the message sequence. They utilize low-rank matrix factorization and simultaneous OMP \cite{Determe2016On} for activity detection and channel estimation, followed by a joint pattern and data detection based on AMP.} Moreover, two slotted pilot-free ODMA schemes are proposed in \cite{Zhang2024Efficient} and \cite{Zhang2024Dictionary}. While an iterative TIN strategy is employed for joint pattern and data detection in \cite{Zhang2024Efficient}, random dictionary learning is studied in \cite{Zhang2024Dictionary} in conjunction with clustering-aided activity detection. \textcolor{black}{Finally, in a recent work \cite{Zhang2025LIM}, the authors present a strategy that is called \textit{Less is More} in a slotted ODMA framework. They propose to transmit a small portion of data using a polar or LDPC code, while embedding the rest of the data into the proposed transmission structure. The transmitted signal consists of a pilot sequence followed by a polar or LDPC codeword. At the receiver side, they employ simultaneous OMP for activity detection followed by zero-forcing symbol estimation, single-user decoding, and SIC. This scheme outperforms the others for fewer than 500 active users, and its performance is slightly better than the achievability bound in \cite{Gao2023Energy} when the number of active users is less than 400. }

\subsubsection{\textbf{\color{black}{Bayesian approach}}}

In pilot-based algorithms, the channel estimation and data detection tasks are performed independently. Hence, the error from one block is propagated to the next one. To resolve this issue, a new set of MIMO URA schemes has been developed that employ a multi-layer iterative Bayesian decoder to jointly estimate each user's transmitted signal and their corresponding channel coefficients \cite{Jiang2023AFully,Han2023Receiver}. Specifically, in \cite{Han2023Receiver}, the message bits are divided into two parts; one part is encoded by a convolutional code, and the other is encoded by index modulation. The transmitted signal is obtained by the Kronecker product of the two components. Additionally, the authors in \cite{Jiang2023AFully} propose encoding the user payload by SPARCs, where all segments of the message bits of a user are mapped onto one codeword. This eliminates the need for an outer tree code, increasing spectral efficiency.

\begin{figure}
	\centering
\includegraphics[width=1\linewidth]{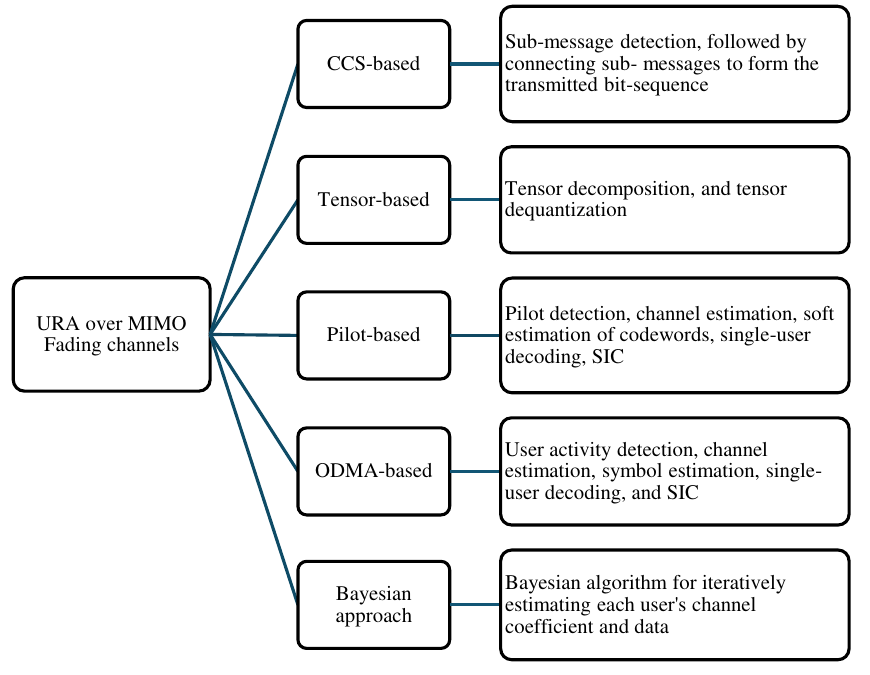}
		\caption{ The decoding steps of the main URA approaches for MIMO fading channels.}	
        \label{UnsourcedStrategies}
\end{figure}

\subsection{\textbf{\textcolor{black}{Summary and Comparison of MIMO Fading Solutions}}}
A summary of MIMO URA schemes is provided in Table \ref{tab_MIMO}, accompanied by the decoding steps of the main approaches summarized in Fig. \ref{UnsourcedStrategies}. 

\subsubsection{\textbf{\textcolor{black}{Performance Comparison}}}
\begin{figure}[t]
	\centering
\includegraphics[trim={.5cm 0cm 1cm 0.6cm},clip,width=1\linewidth]{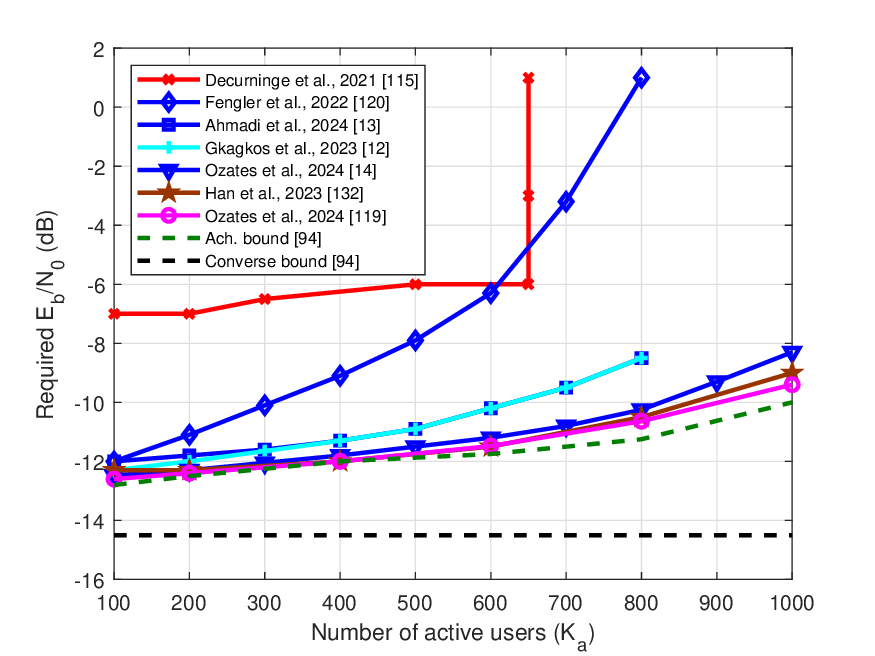}
		\caption{ Comparison of URA schemes for MIMO fading channels with a target PUPE of $0.05$, $n=3200$, and $B=100$ (the common set of parameters in the literature).}	
        \label{FIGMIMO_schemes}
\end{figure}

In Fig. \ref{FIGMIMO_schemes}, a performance comparison of the proposed coding schemes for URA with a massive MIMO receiver is provided for $n = 3200$, $B = 100$, $\epsilon = 0.05$, and $M_r = 50$, assuming that the channel vectors consist of i.i.d. Rayleigh fading coefficients across the different users and different antennas. The results in Fig. \ref{FIGMIMO_schemes} show that, due to the adaptation of strong channel coding algorithms such as LDPC and polar codes, pilot-based schemes in \cite{fengler2022pilot,Gkagkos2023FASURA,Ahmadi2023Unsourced,Ozates2023Aslotted,Ozates2024ODMA} usually outperform the tensor-based scheme in \cite{decurninge2020tensor} for which a relatively weak channel code is utilized, while the SKP coding-based scheme in \cite{Han2023Receiver}, following the Bayesian approach, offers similar performance to the best-performing pilot-based solutions.
\textcolor{black}{Note that, to benefit from increased spatial diversity, the existing URA works on MIMO fading channels assume that the receiver is equipped with a large number of antennas (i.e., the massive MIMO regime), and only a few provide results for a small number of receive antennas \cite{Ozates2023Aslotted}, \cite{Han2023Receiver}.}

\textcolor{black}{Taken together, the MIMO comparisons highlight that leveraging spatial dimensions is critical for scaling URA to 6G-class scenarios, which indicates that MIMO URA is among the most promising paths toward practical deployment, particularly when combined with structured coding and advanced interference management.}


\begin{table*}[ht]
\footnotesize
\centering
\caption{A summary of URA solutions for MIMO fading channels.}
\begin{tabular}{|m{.10\linewidth}|m{.08\linewidth}|m{.47\linewidth}| m{.08\linewidth}|m{.07\linewidth}|m{.05\linewidth}|}
\hline 
\cellcolor[HTML]{BDBDBD}Paper &\cellcolor[HTML]{BDBDBD}Approach& \cellcolor[HTML]{BDBDBD}Description & \cellcolor[HTML]{BDBDBD} \textcolor{black}{Synchron. vs. Asynch.} & \cellcolor[HTML]{BDBDBD} \textcolor{black}{Channel Estimation}  & \cellcolor[HTML]{BDBDBD} \textcolor{black}{Channel Code}
\\\hline
\makecell{Gao et al., \\2023 \cite{Gao2023Energy} }& - & Provides achievability and converse bounds. & \textcolor{black}{-} & \textcolor{black}{-} & \textcolor{black}{-} \\\hline

\makecell{Fengler et al.,\\2021 \cite{fengler2021non}  }& CCS &Concatenated coding scheme where the inner pilots and channel vectors are jointly recovered by ML detection and the recovered segments are stitched using a tree code. & \textcolor{black}{Synchron.} & \textcolor{black}{ML} & \textcolor{black}{-}  \\\hline

\makecell{Shyianov et al.,\\2021 \cite{Shyianov2021Massive}  }& CCS & CCS-based scheme where a variation of AMP is employed for inner CS decoding and channel estimation. The recovered sub-messages are stitched exploiting correlation, which improves the energy efficiency. & \textcolor{black}{Synchron.}  & \textcolor{black}{HyGAMP} & \textcolor{black}{-}  \\\hline

\makecell{Jiang et al.,\\2021 \cite{Jiang2024Two}  }& CCS & A two-stage scheme with a long sub-message in the first stage is used for channel estimation, and very short sub-messages are recovered using orthogonal codebooks in the second stage. & \textcolor{black}{Synchron.}  & \textcolor{black}{AMP} & \textcolor{black}{-}  \\\hline

\makecell{Wang et al.,\\2022 \cite{mimoasync}  }& CCS & A concatenated coding scheme employing SPARCs, tree coding and OFDM. & \textcolor{black}{Asynch.} & \textcolor{black}{Embedded} & \textcolor{black}{-}  \\\hline

\makecell{Decurninge et al., \\2021 \cite{decurninge2020tensor} }& tensor-based& Users' signal separation and detection is done by tensor decomposition. & \textcolor{black}{Synchron.} & \textcolor{black}{Embedded} & \textcolor{black}{BCH} \\\hline

\makecell{Luan et al.,\\2022 \cite{Luan2022Modulation}  }& tensor-based&  A new tensor-based modulation scheme in two versions: one employing QPSK encoding, and the other implementing Grassmann encoding. It shows improved energy efficiency over the tensor-based scheme in \cite{decurninge2020tensor}. & \textcolor{black}{Synchron.} & \textcolor{black}{Embedded} & \textcolor{black}{BCH} \\\hline

\makecell{Fengler et al.,\\ 2022 \cite{fengler2022pilot}  }& pilot-based& A non-slotted scheme employing MMV-AMP pilot detection, LMMSE channel estimation, and single-user decoding. Transmit signal structure: non-orthogonal pilot + polar codeword. & \textcolor{black}{Synchron.} & \textcolor{black}{LMMSE} & \textcolor{black}{Polar} \\\hline
\makecell{ Ahmadi et al.,\\ 2024 \cite{Ahmadi2023Unsourced}  }& pilot-based & A slotted scheme employing NP hypothesis testing pilot detection, MRC channel estimation, single-user decoding, and SIC. Transmit signal structure: Multiple stages of orthogonal pilots + polar codeword. & \textcolor{black}{Synchron.} & \textcolor{black}{MRC} & \textcolor{black}{Polar}
\\\hline
\makecell{ Gkagkos et al.,\\ 2023 \cite{Gkagkos2023FASURA}  }& pilot-based& A non-slotted scheme employing energy test-based pilot detection, LMMSE channel estimation, single-user decoding, and SIC. Transmit signal structure: non-orthogonal pilot + randomly spread polar codeword. & \textcolor{black}{Synchron.} & \textcolor{black}{LMMSE} & \textcolor{black}{Polar}     \\\hline



\makecell{ Ozates et al.,\\ 2024 \cite{Ozates2023Aslotted}  }& pilot-based& A slotted scheme employing gOMP pilot detection, LMMSE channel estimation, single-user decoding, and SIC. Transmit signal structure: non-orthogonal pilot + polar codeword. & \textcolor{black}{Synchron.} & \textcolor{black}{LMMSE} & \textcolor{black}{Polar}
\\\hline

\makecell{ Ozates et al.,\\ 2024 \cite{Ozates2024ODMA}  }& ODMA & A non-slotted scheme employing gOMP pilot detection, LMMSE channel estimation, single-user decoding, and SIC. Transmit signal structure: non-orthogonal pilot + polar codeword spread by ODMA. & \textcolor{black}{Synchron.} & \textcolor{black}{LMMSE} & \textcolor{black}{Polar} \\\hline

\makecell{ Han et al.,\\2023 \cite{Han2023Receiver}  }& Bayesian& SKP coding-based scheme employing  BiG-AMP for outer matrix factorization and BCJR for decoding of the convolutional code. & \textcolor{black}{Synchron.}  & \textcolor{black}{Embedded} & \makecell{\textcolor{black}{Convol-}\\ \textcolor{black}{utional}}  \\\hline
\makecell{ Jiang et al., \\2023 \cite{Jiang2023AFully} }& Bayesian & A Bayesian scheme employing SPARCs for transmission and a three-layer message passing algorithm for decoding. & \textcolor{black}{Synchron.} & \textcolor{black}{Embedded} & \textcolor{black}{-} \\\hline

\makecell{ Li et al., \\2025 \cite{Li2025Asynchronous} }& Bayesian & A Bayesian scheme employing SPARCs, LDPC, and OFDM, where both timing and frequency offsets are considered. & \textcolor{black}{Asynch.} & \textcolor{black}{GB-CR} & \textcolor{black}{LDPC}\\\hline

\end{tabular}
\label{tab_MIMO}
\normalsize
\end{table*}






\subsubsection{\textbf{Computational Complexity \textcolor{black}{Comparison}}}
{\color{black} Comparing the computational complexity of the coding strategies for MIMO URA, we observe that CCS exhibits the lowest complexity among them, as its complexity is linear in $M_r$ and the number of common codebook elements \cite{fengler2021non}. Tensor-based \cite{decurninge2020tensor} and non-slotted pilot-based \cite{Gkagkos2023FASURA} schemes have a similar complexity (cubic in $K_a$). This complexity results from the canonical polyadic decomposition in the former and MMSE filtering for channel estimation in the latter. On the other hand, the complexity of slotted pilot-based schemes \cite{Ahmadi2023Unsourced}, \cite{Ozates2023Aslotted} is cubic in the number of active users in the slot, which demonstrates the complexity advantage of slotting. Furthermore, the complexity of the Bayesian approach in \cite{Han2023Receiver} is linear in $K_a$, $M_r$, and $n$. To summarize, for practical system parameters, CCS has the lowest complexity, while the complexities of slotted pilot-based schemes and the Bayesian approach are similar and smaller than those of tensor-based and non-slotted pilot-based approaches.}

\subsection{\textcolor{black}{Existing URA Solutions on Alternative Setups}}

Most URA schemes in the literature consider direct links between all the users and the BS; however, in certain environments, the direct link between some users and the BS may be blocked or significantly attenuated. In these cases, one solution is to employ RIS, a promising technology that can furnish the URA system with high spectral efficiency and energy savings. Specifically, it has been shown that a passive RIS equipped with many low-cost passive elements, which can intelligently tune the phase-shift of the incident electromagnetic waves, and reflect them in a desired direction without any amplification, can improve the efficiency of a URA system by enabling line-of-sight paths between the transmitters and the receiver in problematic environments with many blocking obstacles \cite{Yan2020Passive,Wu2019Intelligent}. 
More specifically, the RIS-based schemes in \cite{Ahmadi2023RIS,Shao2022reconf} operate in two phases. In the first phase, known as the RIS configuration phase, the BS detects the active pilots and estimates their respective channel coefficients, using which the BS suitably selects the phase shifts of the RIS elements. In the second phase, known as the data phase, the transmitted messages of active users are decoded using SIC-based decoders. 
The RIS-based URA solution in \cite{Ahmadi2023RIS} outperforms the one in \cite{Shao2022reconf}, with a power savings of up to 10 dB. This improvement is due to the employment of a slotted transmission structure, which reduces the interference, as well as a more suitable metric in the RIS design in \cite{Ahmadi2023RIS}, where the overall phase-shift matrix is obtained by minimizing the decoding error probability, in contrast to maximizing the minimum channel gain among active users for the RIS design in \cite{Shao2022reconf}. The authors enhance the RIS design metrics presented in \cite{Ahmadi2023RIS} and further improve upon them in \cite{Ahmadi2024RIS}, obtaining bounds on the system performance.

Conventionally, in MIMO URA systems, ideal hardware units are assumed at both the BS and the UE sides. However, due to the massive number of users and antenna elements, the hardware needs to be inexpensive, and these elements are prone to hardware impairments (HWI) such as I/Q imbalance, phase noise, and quantization errors. The problem of URA with HWI is considered in \cite{hwi}, where the solution developed in \cite{Ozates2023Aslotted} is adapted to the presence of HWI. 

While most URA works consider transmission over i.i.d. Rayleigh block-fading channels, some investigate other channel models. A channel model with one single dominant path between a user and the BS is considered in \cite{Soltani2024Location}, and an approach similar to the one in \cite{nassaji} is proposed, exploiting the user location geometry by linking the transmission features such as the employed preamble or interleaver to the angle of arrival of the user. The dynamics of the user states and channels are taken into account in a slotted structure in \cite{Jiang2024Dynamic}, namely, the users can start or end their transmission in any slot, and they can be active in a subset of the slots, and the channel coefficients change from one slot to another. The authors design a two-step decoding algorithm with an inner AMP and an outer variational message passing. \textcolor{black}{Type-based MA, where each device observes the state of a process and transmits it to a common receiver, is investigated in an unsourced setup in \cite{Ngo2024Type}, where the receiver's goal is to estimate the states and their multiplicity. 
 \textcolor{black}{An achievability bound quantifying the performance of type-based MA is derived in \cite{Krishnan2025Achievability}. Moreover, the authors in \cite{Okumus2025Type} extend the framework in \cite{Ngo2024Type} to the case of fading channels, where a cell-free massive MIMO setup is considered, and a solution exploiting location-based codebook partitioning is proposed.} The problem of user identification and authentication is applied to URA in \cite{Kotaba2021How} with the aid of a message authentication code. Moreover, the authors in \cite{Bayanifar2024Information} propose a URA scheme that combines physical and medium access layers where the medium access layer slotting carries information. The results show that carrying information by the medium access control layer improves the system performance.} A Rician fading channel is considered in \cite{rician} through an application of coded compressed sensing. A correlated fading scenario is tackled in \cite{correlated}. \textcolor{black}{Frequency-selective fading is examined in the context of MIMO URA in \cite{Kharbech2024Massive} and \cite{Liang024Slotted}. The authors in \cite{Kharbech2024Massive} propose a tensor-based modulation scheme, while a slotted scheme employing OFDM is studied in \cite{Liang024Slotted}.}

The asynchronous setup is also receiving increasing attention due to its easier implementation in hardware. See \cite{mimoasync,Musa2023Message,Li2025Asynchronous} for proposed solutions in the MIMO URA setting. A slotted concatenated coding scheme is proposed in \cite{mimoasync} based on SPARCs and tree codes in an OFDM system, where a variation of the ML technique is employed for decoding of the SPARC. The authors in \cite{Li2025Asynchronous} tackle both timing and frequency offsets and propose an asynchronous URA scheme that employs SPARCs, LDPC codes, and multistage orthogonal pilots for transmission. This scheme also incorporates a graph-based channel reconstruction and collision resolution (GB-CR) algorithm for channel estimation and a modified message passing algorithm for decoding. Symbol asynchronicity in URA is investigated in \cite{Musa2023Message}. 

Furthermore, some works have investigated URA in other communication setups such as cell-free MIMO \cite{Gkiouzepi2025Joint, Gkiouzepi2024Step, Hu2024, Zhang2024cellf,Gkagkos2023Scalable,Ozates2025ODMA} and erasure \cite{Zheng2023,Zheng2024} channels. For instance, a cell-free setup is considered in \cite{Gkiouzepi2025Joint} where the coverage area is divided into multiple regions, each served by a receiver unit. Users in each region employ a common codebook, distinct from those in other regions, to transmit their messages. The signals picked up at the receiver units are then jointly processed in a central unit. \textcolor{black}{Joint processing is also considered in \cite{Hu2024} and \cite{Zhang2024cellf}, where a combination of channel coding and fixed sparse coding is employed in the former, while an uncoupled CCS scheme utilizing coarse location information for trellis stitching is proposed in the latter.} \textcolor{black}{On the other hand, \textit{Level 2} cooperation, where only the symbol estimates are passed from the access points to the central processing unit, is considered in \cite{Gkagkos2023Scalable} and \cite{Ozates2025ODMA}. Both schemes are pilot-based, and \cite{Gkagkos2023Scalable} employs the random spreading approach in the data part, while ODMA is utilized in \cite{Ozates2025ODMA}.}

The common assumption in the URA literature is that all users have the same number of bits to transmit or require the same quality of service. However, this may be impractical in situations where the devices have different payload requirements or power budgets, especially considering their massive number. This problem, known as multi-class URA, is addressed in \cite{multi1,multi2} within the coded compressed sensing framework. More specifically, a scenario is considered in which devices are clustered into two classes with different SNR levels and payload requirements. In the cluster with higher power, devices transmit using a two-layer superposition modulation, while those in the other cluster transmit with the same base constellation as in the first cluster. To recover packets at the receiver, a CCS-based approach is employed, where an outer code is used to stitch message fragments together across time and layers.

\begin{figure}[t]
	\centering
\includegraphics[width=1\linewidth]{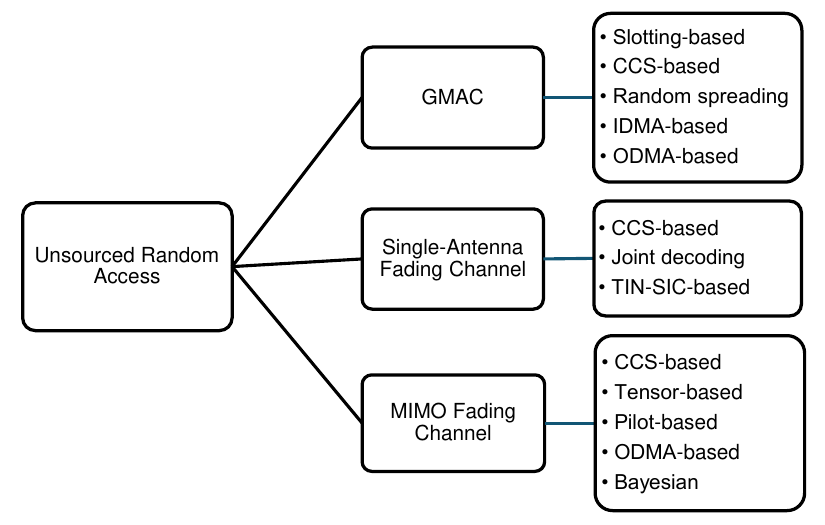}
		\caption{ \textcolor{black}{A summary of different solutions in the URA literature.}}	
        \label{FigURAclass}
\end{figure}

\textcolor{black}{We provide a summary of different URA setups and the corresponding solution approaches in Fig. \ref{FigURAclass}.}

\section{Conclusions}
In this survey, we provided a comprehensive review of the URA literature, with an emphasis on the state-of-the-art. We grouped different URA solutions according to their assumptions on the underlying wireless channels, and explained the representative works in each group in more detail. Moreover, we compared the works in the literature both in terms of setup and performance to gain a clearer understanding of their pros and cons. \textcolor{black}{Across the three major channel models, our survey demonstrates a consistent evolution of URA research: from foundational achievability bounds on the GMAC, to practical yet computationally constrained designs in single-antenna fading channels, and finally to scalable, high-performing solutions enabled by MIMO. This progression reflects the natural trajectory of URA from theory toward real-world adoption in 6G and IoT systems. The comparative analyses highlight that future URA solutions must jointly optimize energy efficiency, interference management, and computational complexity, while also harnessing spatial diversity and structural coding. Taken together, these insights position URA not only as a theoretical construct but as a practical enabler of massive connectivity, capable of meeting the stringent requirements of next-generation networks.}

\subsection{\textcolor{black}{Lessons Learned}}
\textcolor{black}{
The key lessons and insights learned from existing URA solutions can be summarized as follows.}

\begin{itemize}
\item \textcolor{black}{\textbf{ODMA is a promising URA solution.} State-of-the-art URA solutions provide a balance between energy efficiency, scalability, and low-complexity decoding. Among these solutions, ODMA-based schemes excel by offering an effective trade-off between performance, sparsity, and computational complexity. The time-domain sparsity of ODMA-based solutions significantly reduces user interference, making multiuser detection and decoding more manageable even under high user loads, resulting in outstanding performance in both AWGN and fading scenarios, particularly when combined with power diversity. The structured sparsity introduced by ODMA is especially advantageous in massive MIMO systems, where spatial diversity can further enhance separation and decoding accuracy. Additionally, ODMA-based solutions are promising for asynchronous URA, as the ODMA transmission patterns can be utilized instead of commonly employed concatenated preambles to estimate the start times of packets necessary for asynchronous detection, thereby increasing system efficiency. Overall, ODMA-based schemes are promising candidates for scalable, practical, and high-performance solutions for anticipated URA applications, such as massive IoT.}

\item \textcolor{black}{\textbf{Slotted transmission is a simple yet effective interference management technique in URA.} Interference management is one of the major challenges in URA. Among various interference management approaches in the URA literature, ODMA and slotted transmission are promising yet straightforward techniques that introduce structured sparsity by limiting the transmission of user packets to a small portion of the frame, thereby reducing interference. More specifically, by dividing the transmission frame into multiple slots and allowing each user to randomly select one or more slots for transmission, slotting-based solutions decrease the number of simultaneously active users per slot. This localized activity reduces multiuser interference, thereby enabling more accurate signal recovery and effective use of decoding techniques such as SIC. Furthermore, slotted transmission simplifies the decoding process by confining computations to smaller subsets of users, thus decreasing the overall computational complexity compared to full-frame schemes. As a result, slotting serves as a key design element for practical and low-complexity URA implementations.}

\item \textcolor{black}{\textbf{Pilot design and proper channel estimation are critical in URA scenarios over fading channels.} 
In fading environments, reliable channel estimation is essential for accurate user separation and decoding. Since assigning dedicated pilots to each user is infeasible in URA due to scalability constraints, pilot-based solutions often require users to draw pilot sequences from a common, non-orthogonal codebook. This introduces the risk of pilot collisions, which can lead to severe interference and decoding failures. Therefore, minimizing the probability of pilot collisions and optimizing pilot properties, such as length, power, and structure, are crucial for effective activity detection, channel estimation, and interference management. Note that existing pilot-free URA schemes eliminate pilot overhead, but often result in higher complexity and less reliable performance compared to their pilot-based counterparts.}

\item \textcolor{black}{\textbf{Massive MIMO is a game changer in the realization of URA schemes.} Provided with a large number of antennas, the URA receiver can better manage interference and enhance channel estimation accuracy, even under heavy user loads. In particular, massive MIMO enables advanced decoding strategies, such as spatial separation of users and improved compressed sensing techniques, that would be infeasible for a receiver with a limited number of antennas. These capabilities position massive MIMO as a foundational enabler for highly efficient, scalable communication in a URA setup.}

\item \textcolor{black}{\textbf{URA schemes are sensitive to various system parameters.} Certain system parameters can significantly affect the performance, computational complexity, and scalability in URA solutions. Therefore, key parameters, such as the number of active users, frame length, number of slots, pilot sequence length, and power allocation, must be carefully optimized to balance trade-offs between interference mitigation, decoding accuracy, and computational efficiency. Improper tuning of these parameters can lead to severe performance degradation; for instance, too many pilot sequences increase the computational complexity of activity detection, while insufficient pilot diversity raises the risk of collisions. The interdependence of these factors makes robust URA design a delicate task, requiring adaptive parameter configuration based on traffic patterns and deployment scenarios to ensure reliable communication. 
}
\end{itemize}

\subsection{Future Research Directions}
As evident from this survey, numerous works on URA have been conducted in recent years. However, there are still issues, some of which are highly critical, that require further investigation. We will next outline some of these issues and potential methods to address them, and provide several promising future research directions. 

\subsubsection{\textbf{Asynchronous URA} }
    In URA, the common assumption is that the users transmit simultaneously, or more precisely, their signals are received at the same time at the receiver (e.g., symbol synchronicity). However, due to the nature of users and devices in envisioned URA applications, such as massive sensor networks, this transmission coordination is very challenging. This makes asynchronous URA an important and practically relevant research direction to investigate. Note that even if the transmission synchronization is realized, the synchronous reception cannot be guaranteed. Due to the massive number of potential devices with diverse distances to the receiver, the difference in reception times between many users can be larger than a time slot, which makes the setup asynchronous.
    Interestingly, it has been recently shown that if asynchronicity is dealt with properly, it can even lead to performance improvements compared to the synchronous case \cite{Fengler2023}.
    
    There have been some recent attempts in the literature to address asynchronous URA \cite{Kowshik2019sho,andreev2019low,Ozates2023Unsourced,mimoasync,Li2025Asynchronous,Wu2024}. Assuming a bounded time horizon enables these works to convert the problem to one similar to the synchronous scenario, e.g., by employing a limited decoding time frame. For instance, in \cite{Kowshik2019sho}, the authors employ a frequency-domain solution to convert the asynchronous problem to a phase-shift estimation one. Then, conventional synchronous strategies such as TIN-SIC are utilized to decode the active users. 
     On the other hand, \cite{Karami2024} and \cite{Ozates2025Fully} consider a fully asynchronous (i.e., with an unbounded time horizon) setup for URA over GMAC, and a window-based decoding approach.
     

     The primary URA-specific challenge in a fully asynchronous setup for URA over fading channels is the increased computational complexity, which is significantly higher than that of a conventional (non-URA) setup due to the higher number of required preambles.
     More specifically, in an asynchronous GMAC setup, the preambles only serve to identify the start of the packets at the receiver, so even a very small preamble pool would suffice; while in an asynchronous fading setup, they are also needed to estimate the channel coefficients, which are then used for user separation. 
     In a conventional fading setup, the receiver needs to search only over the assigned preambles, whose number equals the number of active users. However, in a URA setup, due to the lack of coordination between the users and the receiver, the number of available preambles must be significantly higher than in the conventional setup; hence, a much higher computational complexity is required for user activity detection and channel estimation. This computational complexity substantially increases in the asynchronous setup with a sliding window receiver, as the receiver shifts and correlates all possible preambles through the window to find the start of the packets as well as the user channel estimates, repeating this process for each shift of the sliding window. This significantly increases the overall computational complexity, rendering the solution practically infeasible. One way to reduce the complexity is to increase the moving steps of the sliding window, at the cost of an overall performance loss. More sophisticated schemes are needed to alleviate this issue.

        \subsubsection{\textbf{Robust URA}} {\color{black} Like any other communication system, the performance of a URA system is affected by the presence of imperfections in the assumed models or employed communication blocks.} For instance, the presence of hardware impairments, e.g., due to non-linearities of the power amplifiers or phase errors in phase-shifters, is highly probable in many practical URA implementations due to cheap and low-precision devices. So far, this challenge is addressed only in \cite{hwi}, where a robust solution is presented for the case of residual hardware impairments in which the remaining hardware impairments after compensation are statistically modeled as additive and multiplicative noises. The analysis of different hardware impairments on the performance of existing URA solutions and providing robust solutions to mitigate them is an interesting research direction.
        
        Examining the URA literature, we can observe that URA schemes are more sensitive to the choice of design parameters than conventional ones. That is, for the URA schemes to achieve acceptable performance, their design parameters must be optimized whenever a system hyperparameter, such as the active user load or the target PUPE, changes. This makes their standardization and practical employment challenging. Therefore, modifications are needed to make the URA solutions robust against changes in system hyperparameters. One approach is to employ stochastic optimization techniques, such as optimizing the expected value of the cost metric (e.g., PUPE) over the statistics of the intended system hyperparameters, to obtain a set of design parameters that perform well across a wide range of system hyperparameters.

\subsubsection{\textbf{Machine Learning (ML) for and with URA} }
Current solutions look at the URA problem in a single shot. However, the transmission is recurrent and spans several frames in many practical applications. In such cases, ML techniques, such as reinforcement learning, can be employed to adapt to the environment and learn suitable transmission strategies. ML techniques can also be used to simplify certain blocks in the decoding process, such as user activity detection (similarly to the deep learning-based solutions in \cite{deSouza2023}) and channel estimation. For instance, in \cite{Zhang2024}, the authors demonstrate that employing a dictionary learning-based solution eliminates the need for pilots for channel estimation in URA scenarios over fading channels while maintaining an acceptable performance.

Apart from general challenges in implementing ML solutions in communication systems, such as energy and computational constraints and compatibility with legacy systems, there are some that are specific or more pronounced in a URA setup, which is mostly due to the massive number of potential users (up to the order of millions) and their sporadic activity. Training a neural network with an input of very large size (due to the massive number of potential users) with non-stationary characteristics (due to the sporadic activity) is a very challenging task.


\textcolor{black}{URA can also be employed to enable distributed learning among wireless agents. In federated learning (FL), the goal of the parameter server, which acts as the orchestrator, is to aggregate local model updates from many clients at each iteration, and update the global model by averaging these models \cite{Chen2021}. When clients are co-located, and the parameter server is a wireless access point, model updates have to be communicated over a shared wireless channel. In conventional methods, this requires sharing the available channel resources among the active clients; however, in FL, the parameter server is not interested in individual updates. Therefore, in \cite{Amiri2020}, over-the-air computing (OAC) was proposed to exploit the signal superposition property of wireless signals. This would allow the parameter server to directly receive the summation of the model updates.} 

\textcolor{black}{However, such an analog over-the-air computing scheme would require a certain level of coordination and synchronization among the clients. An alternative digital OAC scheme is proposed in \cite{Qiao2024} using URA. In OAC with URA, each client quantizes its model update, and sends the corresponding codeword. Both the quantization and channel codebooks are shared among all the clients. Differently from conventional URA, which relies on the fact that active users are likely to transmit independent codewords, and the receiver tries to identify only the transmitted codewords by the active clients, here the users are more likely to transmit the same codeword as their updates tend to converge as the training process progresses. Moreover, the receiver wants to identify not only the active codewords, but also how many clients transmit each codeword. This approach can be extended to other functions, and the identification of the optimal quantization and communication codebooks is an open research direction. } 


\subsubsection{\textcolor{black}{\textbf{Semantic and Task-Oriented URA}}} 
\textcolor{black}{Conventional URA frameworks are designed under the assumption that all transmitted bits are equally important, focusing primarily on reliable message recovery. However, in many IoT and machine-type communication scenarios, the goal is not to recover raw data streams but rather to support specific tasks such as anomaly detection, control signaling, or status monitoring. In such cases, transmitting semantic or task-oriented information, i.e., only the most relevant features of the data, can drastically reduce the required payload and improve spectral efficiency. Future research may investigate how semantic encoding techniques can be combined with URA to enable resource-efficient access, for example, by allowing multiple devices reporting correlated measurements to share compressed semantic representations. Such integration would align URA more closely with emerging 6G paradigms, where communication systems are expected to deliver knowledge and actionable insights rather than just data.}

\subsubsection{\textbf{ISAC in URA}}
Existing research on URA has focused primarily on the transmission of information from communication users. However, with the evolution of 6G and beyond, new functionalities such as sensing, computing, and security are becoming critical alongside communication. As the hardware architectures and signal processing methods for these tasks are mostly similar, jointly adopting them significantly reduces hardware costs compared to separate systems while maintaining comparable performance \cite{Nikbakht2024}. A promising example of such joint designs is integrated sensing and communication (ISAC) systems.

The problem of ISAC in URA is introduced for the first time in \cite{Ahmadi2024Integrated}, where a massive number of unsourced and sporadically active users are connected to the system to jointly perform sensing and communication tasks. In the resulting system, called unsourced ISAC (UNISAC), users are divided into two main categories: Sensing and communication users. The system aims to decode signals from communication users while detecting signals from active sensing users and estimating their AOAs. The authors derive an achievability bound for UNISAC to demonstrate its effectiveness in enabling both user sensing and user communication within a URA setup. This work is conducted under two key assumptions: the targets are active, and the channels between the users (targets) and the BS are line-of-sight (LoS).

The problem of UNISAC has been further investigated in subsequent works, including: proposing a practical transceiver scheme for UNISAC under LoS channels between the BS and all sensing and communication users~\cite{Ahmadi2025Practical,Zhang2025TwoPhase}, as well as under LoS target–BS channels and Rayleigh fading user–BS channels~\cite{Zhang2025Integrated}; studying the UNISAC model equipped with fluid antenna systems~\cite{Zhang2025OnFund,ZhangLWC2025}; and addressing the problem of passive target sensing~\cite{Ahmadi2025Feedback}.

There are several promising directions for future research. In existing UNISAC setups, only the problem of user angle estimation has been investigated, whereas the field of sensing is much broader. For example, tracking, surveillance, passive source localization, and target localization are important sensing tasks that warrant further study. Moreover, networks incorporating advanced technologies, such as multi-BS systems, UAV-equipped systems, and RIS-aided systems, can be explored to enhance performance. Another significant research direction is to study UNISAC under realistic channel models, including frequency-selective fading, large-scale fading, Doppler effects, asynchronous transmission errors, and hardware impairments \cite{Ahmadi2024TWCIntegrated}.

To explore the challenges that future work on URA-based ISAC may encounter, many aspects still require further investigation. In a URA system, a massive number of users transmit sporadically without being registered or authenticated by the network. Therefore, there is no control over transmitter characteristics such as location, signal structure, and beam direction, which are critical when the transmitters' echoes are used for reliable passive target sensing. Additionally, the random access nature of user transmissions causes heavy interference, a phenomenon uncommon in traditional sensing techniques. Advanced algorithms are necessary to address these challenges and enhance the system’s ability to effectively sense passive targets.

    \subsubsection{\textbf{URA with Feedback}}
    {\color{black} In communication systems, feedback has been shown to greatly simplify transmission and reception mechanisms and, in some cases, even improve system performance \cite{Love2008}. The main challenge in employing feedback-based techniques in URA is the lack of coordination between the users and the receiver due to the massive number of potential users.} However, even in URA, there can be some level of coordination as long as the signal sent from the receiver is not user-specific, as there is no user identity in URA. For instance, the receiver can broadcast a common pilot or beacon to the users, allowing them to check if they are among the successfully decoded users from the previous round and estimate their channel coefficients or at least their large-scale components, depending on their dedicated time slots. Note that the time dedicated to this operation needs to be considered as extra overhead when comparing with solutions without feedback. This issue is tackled in part in \cite{Ahmadi2025Feedback,Ebert2022,Bashir2023,Che2024}. In \cite{Ebert2022}, hashes of the messages of successfully decoded users and their estimated channel coefficients are fed back using beamforming techniques. In \cite{Ahmadi2025Feedback}, the authors enhance the approach of \cite{Ebert2022} by informing both the successfully decoded users and those that were not decoded about their respective transmission outcomes. In \cite{Bashir2023}, with the help of a designed feedback beacon, the users estimate their channels and infer whether they were successfully decoded using a predefined threshold. An application of URA can be in sensor networks where the devices partially observe a common state, which is the main interest of the receiver. For such an application, the authors in \cite{Che2024} devise a mechanism where the receiver broadcasts a reliability indicator of the estimated state after each transmission, using which each device can decide whether to transmit in the next iteration.
    
    Recall that the main novelty of URA is providing a unified framework to decouple user identification and data transmission problems. This makes it a suitable approach in applications where user identification is not feasible due to the massive number of potential users. 
    However, URA remains applicable and is a good candidate in non-massive scenarios with a considerable number of potential users, where user identification is still feasible. In these scenarios, conventional authentication and acknowledgment schemes impose a huge overhead on the system. This issue is tackled in \cite{Kotaba2021}.

    \subsubsection{\textbf{RIS-Aided URA}} In URA, due to the massive number of potential users, it is very probable that the line-of-sight link between some users and the receiver is either not present or heavily attenuated. In this case, employing relaying techniques such as RIS can play a crucial role in enhancing the overall system performance by providing an alternative communication link with potentially many degrees of freedom. {\color{black} However, the huge channel estimation overhead is the main challenge in the use of RIS in URA systems.} The channel estimation overhead is already one of the main issues in RIS systems due to the high number of channel coefficients per user as a result of the large number of RIS elements and the potentially massive number of antennas at the receiver (as the benefits of a RIS system are amplified when combined with a massive MIMO receiver). This problem is more pronounced in a RIS-aided URA system as there is a potentially massive number of even users, which renders most existing channel estimation schemes in RIS infeasible. {\color{black} There are a few works that have considered the RIS-aided URA scenario \cite{Ahmadi2023RIS,Shao2022reconf,Ahmadi2024RIS}. Although these works make channel estimation feasible in a RIS-aided URA setup employing an iterative SIC-based approach and time frame slotting strategies, the high computational complexity remains a significant challenge.} 

    Moreover, there are many promising RIS configurations and types that need to be investigated in a URA setup, such as active or hybrid RIS \cite{Yildirim2021} and metasurface-based RIS \cite{Gros2021}, in order to find suitable candidates. Moreover, the employment of massive arrays such as extra-large massive MIMO and RIS leads to some active users being in the near-field of these arrays, which makes hybrid near-field-far-field analysis necessary. This is in contrast with the existing URA solutions, which are all for the far-field regime. 

\subsubsection{\textbf{\textcolor{black}{Security}}}
\textcolor{black}{
One of the key concerns in multiple access systems is a robust implementation against security threats, such as jamming and eavesdropping. To address these security threats, some common countermeasures are employed, including \textit{mutual authentication} protocols for access control and authentication, such as the authentication and key agreement (AKA) used in 4G/5G, and data encryption techniques, such as AES-128 in 4G and AEGIS-256 in 5G. These protocols ensure that both users and the receiver (network) verify each other's identities before establishing a secure connection.
However, many of these solutions are not feasible in a URA setup, primarily due to the lack of coordination between users and the receiver. For example, key agreement in the AKA protocol relies on some level of coordination between the user and the receiver to establish a secure connection.}

\textcolor{black}{To the best of our knowledge, there has not been any work addressing the security aspects of URA. However, some existing security solutions can still be utilized in URA with certain modifications or considerations. For example, light encryption methods can be employed (similarly to those in \cite{Ma2019} and \cite{Esfahani2019}) to introduce a certain level of security with a common key among users. In some URA solutions, the key can also be randomly selected from a pool, with each key paired with a transmission signature, such as a pilot sequence in pilot-based schemes and a transmission pattern in ODMA-based ones. Likewise, physical layer security techniques, such as the injection of artificial noise, can also be employed, with the modification that all users utilize the same noise sequence or randomly select it from a pool of available sequences. Existing physical layer authentication techniques based on signal features, such as channel state information or RF fingerprints \cite{Ureten2007}, can also be utilized as long as they do not rely on user-specific signatures.
}

    \subsubsection{\textbf{Standardization}}
    There has been no effort in standardizing URA so far.
    This requires a thorough investigation and analysis of the current competing solutions, and finding a good candidate that can preferably support both the current protocols, perhaps with minor modifications, and the high demands of the future ones. For instance, in \cite{Agostini2024}, with a slight modification in the 5G new radio (5G-NR) release 16 protocol, the authors achieve an improvement in energy efficiency of about 5 to 10 dB compared to the existing 5G-NR solution for systems with a small number of active users (10 to 50 users), employing an IDMA-based URA solution. However, this solution is not scalable to higher active user loads.
    
    There are two main challenges in incorporating URA in current protocols, such as 5G-NR and beyond, for reasonable active user loads (e.g., in the order of hundreds). The first is the small number of preamble sequences available in 5G-NR as opposed to the massive number of preamble sequences (pilots) needed in the preamble/pilot-based URA solutions. The second is the limited number of access patterns that can be realized, which is vital in transmission pattern-based URA solutions such as ODMA-based ones. This indicates that a crucial step in realizing URA in future protocols is to considerably increase the available pilot or transmission patterns.

This comprehensive survey of URA reveals an active field that has evolved significantly in recent years, with substantial progress in fading and MIMO channels. The research challenges identified above demonstrate both the complexity of remaining questions and the rich opportunities for future investigation. Addressing these challenges will be essential for the practical adoption and deployment of URA solutions in future wireless systems.

\end{document}